\documentclass[a4paper,11pt]{article}
\usepackage{amsthm,amsmath,amssymb,amsfonts}
\usepackage{graphicx}
\usepackage{appendix}
\usepackage{color}
\usepackage{times}
\usepackage{bm}
\usepackage{natbib}
\usepackage[plain,noend]{algorithm2e}

\usepackage{multirow}

\def\T{{ \mathrm{\scriptscriptstyle T} }}

\newcommand{\R}{\mathbb{R}}

\newcommand{\intA}{\nu_A}
\newcommand{\intB}{\nu_B}

\newtheorem{theorem}{Theorem}[section]

\newtheorem{remark}[theorem]{Remark}
\renewcommand{\P}{P}
\newcommand{\E}{E}

\newcommand{\telque}{\,:\,}
\newcommand{\FDR}{\mbox{FDR}}

\newcommand{\FWER}{\mbox{FWER}}

\newcommand{\mtc}{\mathcal}
\newcommand{\mbf}{\mathbf}

\newcommand{\wh}[1]{{\widehat{#1}}}

\newcommand{\ind}[1]{\mbf{1}_{\left\{#1\right\}}} %{\mbf{1}_{\left\{#1\right\}}}

\newcommand{\eps}{\varepsilon}

\renewcommand{\l}{\ell}

\newcommand{\pvalue}{$p$-value$\,$}
\newcommand{\pvalues}{$p$-values$\,$}
\newcommand{\minp}{$\min$-$p\,$}
\newcommand{\Nsf}{\mathfrak{N}}
\newcommand{\lab}{\lambda}
\newcommand{\m}{\text{m}}

\begin{document}

\title{Continuous testing for Poisson process intensities :\\ A new perspective on scanning statistics}
\author{Franck Picard$^\star$\\
$^\star$ \textit{Univ Lyon, Universit\'e Lyon 1, CNRS, LBBE UMR5558},  \\ {F-69622 Villeurbanne, France} \\
franck.picard@univ-lyon1.fr\\
\\
Patricia Reynaud-Bouret$^\ddag$\\
$^\ddag$ \textit{Universit\'e C\^ote d'Azur, CNRS, LJAD, France},\\
Patricia.Reynaud-Bouret@unice.fr\\
\\
Etienne Roquain$^\dag$\\
$^\dag$  \textit{LPMA, UMR 7599, Sorbonne Universit\'es, UPMC, Univ. Paris 6},\\  {F-75005, Paris, France.}\\
etienne.roquain@upmc.fr 
}

\maketitle
\date{}

\begin{abstract}
We propose a novel continuous testing framework to test the intensities of Poisson Processes. This framework allows a rigorous definition of the complete testing procedure, from an infinite number of hypothesis to joint error rates. Our work extends traditional procedures based on scanning windows, by controlling the family-wise error rate and the false discovery rate in a non-asymptotic manner and in a continuous way. The decision rule is based on a \pvalue process that can be estimated by a Monte-Carlo procedure. We also propose new test statistics based on kernels. Our method is applied in Neurosciences and Genomics through the standard test of homogeneity, and the two-sample test.
\end{abstract}

\newpage
\tableofcontents
\newpage

\section{Introduction}

Continuous testing has recently emerged as a suitable theoretical framework to test a possibly infinite number of hypothesis. It is especially adapted to situations where observed data come from an underlying continuously-indexed random process, like in the classical white noise model \citep{DS01}, or when data can be modeled by a point process \citep{BDR2014}. Our work being motivated by data coming from Neurosciences and Genomics, we focus on point processes that have been very effective to model occurrences of events through time or space in many fields. More than a broad range of applications, point processes have the advantage to be discrete by essence, which allows a drastic reduction of the continuum number of hypotheses, into to a random finite number of tests. This property was exploited by \cite{BDR2014} in a particular context, and we elaborate upon this work to propose a continuous testing framework that allows the rigorous definition of the complete testing procedure, from multiple hypothesis to error rates. Our procedure is very general and fully operational thanks to the use of sliding windows. This allows us to we revisit the old-tale of scanning statistics, and we extend their traditional possibilities and performance by providing the theoretical controls for the Family Wise Error Rate (FWER) as well as for the False Discovery Rate (FDR). We focus on Poisson processes for their remarkable properties, but the main definitions, methodological developments or computational tricks are valid for general point processes.

Scanning windows probably constitute the most common method to perform tests on Poisson processes for the detection of unexpected clusters of observations \citep{K97,CZ07,PGVW2007,SZY2011} or to compare samples \citep{W10}. It consists in computing a test statistic on overlapping windows of a given length that are shifted by all possible delays, which has the advantage of avoiding empirical discretization of the data \citep{grunt,MTGAUE}. Then the decision rule is based on the so-called scan statistic, that refers to the window with most signal. Therefore these procedures rely on the distribution of the maximum of the test statistics under the full null, that can be either approximated \citep{Nau1982,RRS2005,FWL2012,RW13,W10,CZ07} or learned by a randomization procedure \citep{KHH05,KHK09,ACT15}, which is common in standard multiple testing procedures \citep{WY1993,RW2005}. In this work we adopt the conditional testing approach \citep{L91,RW13} that has the advantage of getting rid of nuisance parameters. It consists in deriving tests that are conditional to the \textit{number} of occurrence (for the homogeneity test) or to the \textit{position} of occurrences (for the two-sample test). Also, this strategy is consistent with the strong connections between continuous testing and non-parametric statistics \citep{BDR2014}, since the conditional test may depend on the unknown intensity functions of the observed processes. Therefore, our strategy is also to go beyond standard parametric approaches, that are mainly based on observed counts or on a likelihood ratio statistic. As we will show, these parametric statistics do not fully exploit the spatial/temporal information that lie within the data, but only use integrated information. Inspired by recent results on non-parametric testing \citep{GBR12,FLRB13}, we enrich the scan-statistics toolbox with kernel-based statistics to increase the power of detection of sharp alternatives.

Another limitation of current window-based strategies is that they correspond to global procedures that focus on a global null hypothesis. This resumes to ask whether the process is globally homogeneous, or if the samples are globally identical, which inevitably results in a global yes/no answer. However, with the increasing availability of more and more complex data, there is a need for flexible and local approaches to improve sensitivity for local structures. As we will see, continuous testing allows us to revisit the scanning windows framework, by defining a (possibly) uncountably infinite set of local hypothesis, and most importantly by providing a rigorous framework for defining joint error rates.

Indeed, the well known statistical challenge that arises when dealing with sliding windows is the appropriate control of Type-I errors, which is made difficult by the induced dependency structure. Current strategies mostly rely on the asymptotic approximation of the scan distribution \citep{K97,CZ07}, which has two advantages : $i)$ it somewhat avoids the dependency matter by focusing on the scan statistic only, $ii)$ it is computationally efficient. We rather adopt a non-asymptotic point of view, by considering the complete \pvalue process that becomes here a stochastic process . For some particular statistics (the count statistic for instance), this process is easily computed. When the statistic distribution is unknown we propose a Monte-Carlo procedure to estimate the \pvalue process. Then we propose a \minp procedure in continuous time to control the FWER in a non-asymptotic manner, based on a new computationally efficient double Monte-Carlo scheme. Simulations show that this non asymptotic procedure outperforms the one of \cite{CZ07} in terms of FWER control.

The control of the False Discovery Rate (FDR), that is the expected proportion of errors among rejections, is more intricate, and tentatives are more recent. \cite{PGVW2004} propose a continuous analogue to the FDR in the context of random fields, where the amount of null hypotheses is assessed by the Lebesgue measure. However, their (continuous) FDR controlling procedure is based upon a FWER guarantee. Their strategy has also been considered in a setting close to our homogeneity framework \citep{PGVW2007}. \cite{SZY2011} also investigated the FDR for scanning statistics, but only under two strong conditions (Poisson distribution for the number of false discoveries and independence between null and alternative statistics), none of which being satisfied in standard models. Moreover, their procedure eventually resumes to discretized \pvalues, and is based on the tail approximation of the scan statistic, which is not valid for other statistics other than the scan statistic. Thanks to the continuous testing framework, we propose a weighted Benjamini-Hochberg procedure in continuous time for Poisson processes testing, which outperforms on simulations \cite{SZY2011}. This procedure corresponds to a continuous analogue of the well-known step-up procedure of \cite{BH1995}, and extends the results of \cite{BDR2014} for continuous FDR in the context of local hypothesis testing.

The present work is inspired by two applications of Poisson processes: the analysis of spike trains data in Neurosciences and analysis of Genomic data coming from Next Generation Sequencing (NGS) technologies. In Neurosciences, Poisson processes are used to model spike trains, that is the succession of action potentials of a given neuron \citep{Pipa2013}. In particular inhomogeneous Poisson processes constitute a powerful model to understand rapid changes in firing rates induced by some stimulii \citep{KassVenturaCai}. While the estimation of the Poisson process intensity has been already deeply studied \citep{SS10}, the questions asked in practice are often somehow different from a pure curve reconstruction matter: does the stimulus affect the spike apparition process (global hypothesis)? When (local) ? Does it depend on the type of stimulus and if so when exactly are the differences? In Genomics, next generation sequencing technologies (NGS) now allow the fine mapping of genomic features along chromosomes, and this spatial organization has become central to better understand genomes architecture and regulation. Here we consider the spatial organization of human replication origins that constitute the starting points of chromosomes duplication. To maintain the stability and integrity of genomes, replication origins are under a very strict spatio-temporal control, and part of their positioning has been shown to be associated with cell differentiation \citep{PCA14}. Their recent mapping on the human genome has allowed the comparative analysis of replication origins maps between cell lines. Unfortunately, there is a lack of statistical framework to compare genomic maps of spatially organized genomic features, and we show how our procedure can be applied to the identification of replication starting sites that are linked to cell differentiation.

This article is organized as follows. First, we define the continuous testing framework, and we show how it can be easily handled in the Poisson process set-up (Section \ref{setup}), with a particular focus on the test of homogeneity and the two-sample test. We also define the infinite set of local test hypotheses, the continuum of scanning windows to test them and the associated \pvalue stochastic process. To proceed, and for the sake of simplicity, we first focus on the standard count statistic that directly gives access to the \pvalues. Then two types of control are discussed in Section \ref{control}: the classical FWER and the more involved control of the FDR in continuous time. Both procedures result in an adjusted \pvalue process, which makes our method operational and easily implemented. To increase the power of our procedure we propose in Section \ref{kernel} local kernel-based statistics. Those kernel statistics require non explicit conditional distributions to be transformed into \pvalues and we discuss Monte-Carlo procedures to estimate the \pvalues, with a new computationally efficient double Monte-Carlo step to control the FWER in Section \ref{boot}. In Section \ref{sim} we show through simulations that the resulting procedures are more powerful than existing ones (in particular those based on unconditioned scanning statistics or on discretization). Section \ref{real} shows that the procedure works extremely well on experimental data, issued from Neurosciences or Genomics. Our method is available in the form of the \texttt{contest} \texttt{R}-package, and offers many perspectives in the field of continuous testing.

\section{\label{setup} The continuous testing framework for Poisson Processes}

In the sequel a point process is a random countable set of points usually denoted by $N$. The corresponding point measure is then denoted by $dN$, that is $\sum_{T\in N} \delta_T$ with $\delta$ denoting the Dirac mass. For any interval $I$, $N(I)$ is a random variable that corresponds to the number of points of $N$ in $I$. We focus on processes in $\R$ (which can be thought as the DNA strand for genomic applications and time line for neuroscience applications) and  we consider processes in [0,1] for the sake of simplicity.

\subsection{Homogeneity and two-sample conditional tests}\label{frame}

In order to propose a unified framework, we use the following notations: we observe a random set of points $Y$ with a distribution of the form $P_{\theta,\lab}$, where  $\lab$ is a nuisance parameter, while  $\theta$ represents the signal of interest. In both cases, the aim is to test whether $\theta$ is locally null.

\paragraph{Homogeneity test.} We observe $Y=N$, a point process modeled by a heterogeneous Poisson Process on $[0,1]$ with unknown intensity $\nu$ (non negative and in $L^2[0,1]$) with respect to the Lebesgue measure. We would like to detect the regions that are particularly rich or poor in occurrences of the process, with respect to its mean/stationary behavior. To do so, we compare the intensity $\nu$ of the observed process to the mean behavior of the intensity 
$$\lab=\int_0^1 \nu(s)ds.$$ 
Then we rewrite the distribution of $Y$ in terms of two different parameters $\lab$ and $\theta$, with  $\theta$, defined by 
$$\forall t \in [0,1],\, \theta(t) = \frac{\nu(t)}{\int_0^1 \nu(s)ds} - 1.$$ 
The signal of interest is $\theta$, whereas $\lab$ may be seen as a nuisance parameter because its value is unknown under the null hypothesis. The homogeneity null hypothesis is therefore reduced to "$\theta=0$": it is a composite hypothesis since under the null hypothesis, the distribution can still vary according to $\lab$. As a particular easier problem, one can also assume that $\lab$ is known. The problem is then a classical goodness-of-fit test and the null hypothesis is not composite anymore.

\paragraph{Two-sample test.} We observe here $Y=(N_A,N_B)$, where $N_A$ and $N_B$ are two heterogeneous independent Poisson Processes on $[0,1]$ with intensities $\intA$ and $\intB$ (non negative and in $L^2[0,1]$) with respect to the Lebesgue measure. In the following, for any interval $I$, we denote $Y(I)=(N_A(I),N_B(I))$, and by a slight abuse of notation, $(Y \cap I) = \big(N_A\cap I,N_B\cap I\big)$. Our aim here is to detect the regions where $\intA\not = \intB$ (or, alternatively, $\intA>\intB$). 

To re-parametrize the distribution of $Y$, one can provide a one-to-one correspondence between $Y$ and the couple $(N,\eps)$, where $N=N_A\cup N_B$ is the joint process and where $\eps=(\eps_T)_{T\in N}$ is a set of marks with values in $\{-1,1\}$ such that $\eps_T=1$ if $T$ actually belongs to $N_A$ and $\eps_T=-1$ if $T$ actually belongs to $N_B$ \citep{K93}. An important point here is to note that $N$ is actually a Poisson process with intensity $\lab=\intA+\intB$ and that conditionally to $N$, the $\eps_T$'s are independent variables with distribution
\begin{equation*} \label{theta}
\eps_T |  N \sim 2\mathcal{B}\left( \frac{\theta(T)+1}{2} \right)-1,
\end{equation*}
with
$$
\forall t \in [0,1],\, \theta(t) = \frac{\intA(t)-\intB(t)}{\intA(t)+\intB(t)}.
$$
Once again, $\theta$ stands for the amount of signal in the data and $\lab$ is a nuisance parameter. The two-sample null hypothesis is then once again summarized by "$\theta=0$". Since $\lab$ is an unknown function, this hypothesis is composite. This case is thus much more complex than the homogeneity case, and calls for a non-parametric procedure.

\paragraph{Conditional distributions.}  To avoid in both cases the use of specific unknown values of $\lab$ to derive the distribution under the null hypothesis, our strategy is based on conditional testing \citep{L91,RW13}. Indeed, for the homogeneity test, conditionally on the total number of points, $N([0,1])=n$, the observed process is actually an $n$ \textit{i.i.d.} sample of density $1+\theta$ that does not depend on $\lab$ \citep{L91,RW13}. Similarly, in the  two-sample case, given the positions of the joint process $N$, the remaining variability under the null hypothesis lies in the distribution of the marks $(\eps_T)_{T\in N}$ that only depends on $\theta$ (and not on $\lab$ anymore). Thus we introduce the variable  $\Nsf$, such that $\Nsf = N([0,1])$ in the homogeneity case, and $\Nsf = N$ in the two-sample case and our testing procedure relies on the distribution of $Y$ conditionally to $\Nsf$. Since the conditional distribution of $Y$ given $\Nsf$ does not depend on $\lab$, it is denoted by $P_\theta$ in the sequel, whereas $\E_\lab$ refers to the expectation w.r.t. $\Nsf$.

\subsection{An infinite set of local null hypothesis}

The continuous testing procedure developed here aims at determining where the signal $\theta$ is non zero. The classical strategy consists in testing the  \textit{full} null hypothesis, which corresponds to the single question ``Is the signal $\theta$ null over [0,1]?''. Therefore a unique hypothesis is tested and the test produces a binary answer. Here, we rather consider local null hypothesis such as,
\begin{eqnarray*}
\forall t \in [0,1], \,\, H_{0,t}: \big\{\theta(t)=0\big\} &\text{ against } H_{1,t}:& \big\{\theta(t) >0\big\},\label{testproblem1}\\
\text{or } \forall t \in [0,1], \,\,H_{0,t}: \big\{\theta(t)=0\big\} &\text{ against } H_{1,t}:& \big\{\theta(t) \neq 0\big\}\label{testproblem2}.
\end{eqnarray*}
In practice, one-sided tests of homogeneity are usually considered to detect regions that are too rich in occurrences, whereas two-sided tests of the two-sample problem seem more relevant when there is a complete equivalence between both processes $N_A$ and $N_B$ from a modeling point of view. 

Since information at the single point level seems difficult to catch in general, we focus on finding intervals or unions of intervals on which the null hypothesis holds. We denote $\mathcal{H}_0\left\{I\right\}$, the null hypothesis corresponding to $``\forall t \in I, \theta(t)=0"$. As a particular case,  $\mathcal{H}_0\left\{[0,1]\right\}$ is the full null hypothesis and corresponds to $"\theta=\theta_0"$, where $\theta_0$ is  the null function on $[0,1]$. 

\subsection{A continuum of scanning windows}\label{sec:scanwindows}

Our procedure is based on scanning windows with a given resolution $\eta \in ]0,1[$ that is fixed. We envision here all the possible {\it continuum} of windows with such given length, meaning that we are performing a whole {\it continuum} of tests, one for each window. To avoid any confusion in the sequel, $x$ always denotes a \textit{window center} whereas $t$ always denotes a \textit{point}, that is a possible value for the observations $Y$. Thus we have $x \in \mathcal{X}_\eta=[\eta/2,1-\eta/2]$ (while the set of possible values for $t$ is $[0,1]$). The scanning windows are denoted by $I_\eta(x)=(x-\eta/2, x+\eta/2]$ for any $x\in\mathcal{X}_\eta$. Therefore, a typical relationship between  $x$ and  $t$ is $"t\in I_\eta(x)"$. In the sequel, all the proposed multiple testing procedures are actually based on single tests of the null hypothesis $\mathcal{H}_0\left\{I_\eta(x)\right\}$, for all the possible window centers $x\in \mathcal{X}_\eta$, which is different than the null hypotheses $H_{0,t}$ given by $\theta(t)=0$ for $t$ in $[0,1]$. The continuous testing procedure aims at accepting the set of "true windows", which is indexed by their centers:
$$
J_0^\eta(\theta) = \big\{x \in \mathcal{X}_\eta \::\:   \forall t \in I_\eta(x), \theta(t)=0\big\}.
$$
$J_0^\eta(\theta)$ formally depends on $\theta$, but is denoted by $J_0^\eta$ in the sequel,  for short.

\subsection{A first simple construction of the \pvalue process}

\paragraph{Homogeneity test.} Consider as an individual test statistic $S_\eta(x)=N(I_\eta(x))$,that is the number of points in the window of center $x$. Then classically,  a one-sided test based on this statistic rejects $\mathcal{H}_0\left\{I_\eta(x)\right\}$ if  $S_\eta(x)$ is large enough. Under $\mathcal{H}_0\left\{I_\eta(x)\right\}$, it is easy to see that the conditional distribution of $S_\eta(x)$ given $\Nsf=n$ (or $N([0,1])=n$ here) is a Binomial variable with parameter $n$ and $1/2$. Hence, if we denote by $F_{\mathcal{B}(n,\frac{1}{2})}(.)$, the function such that $F_{\mathcal{B}(n,\frac{1}{2})}(z) = P(Z \geq z)$ with $Z \sim \mathcal{B}(n,\frac{1}{2})$, then 
$$\forall x \in \mathcal{X}_\eta,\,\,  p_\eta(x) = F_{\mathcal{B}(N([0,1]),\frac{1}{2})}(S_\eta(x)),$$ can be seen as a \pvalue associated to the classical one-sided single test  of $\mathcal{H}_0\left\{I_\eta(x)\right\}$. Since $F_{\mathcal{B}(n,\frac{1}{2})}$ is explicit and since the number of points, as a function of $x$, can be very efficiently computed, one can have access to a \pvalue process $x\mapsto p_\eta(x)$, with a very low computational cost.

If $\lab$ is known, one does not need to use a conditional distribution. The unconditional distribution of $S_\eta(x)$ is a Poisson distribution with mean $\eta\lab$, under $\mathcal{H}_0\left\{I_\eta(x)\right\}$. If $F_{\mathcal{P}(\eta\lab)}(z)=P(Z\geq z)$ with $Z \sim \mathcal{P}(\eta \lambda)$, then
\begin{equation}
\label{Eq:pvHom}
\forall x \in \mathcal{X}_\eta,\,\, p_\eta(x) = F_{\mathcal{P}(\eta\lab)}(S_\eta(x))
\end{equation}
also defines a \pvalue process. Note that this set-up is quite close to the one of \cite{CZ07}.

\paragraph{Two-sample test.} Consider now
as individual test statistic 
$S_\eta(x)=N_A(I_\eta(x)).$
A classical one-sided test based on this statistic also rejects $\mathcal{H}_0\left\{I_\eta(x)\right\}$, if  $S_\eta(x)$ is large enough. Under $\mathcal{H}_0\left\{I_\eta(x)\right\}$, it is easy to see that the conditional distribution of $S_\eta(x)$ given $\Nsf=N$ (the joint process here), is the same as the conditional distribution of $S_\eta(x)$ given $N(I_\eta(x)) = N_A(I_\eta(x))+N_B(I_\eta(x))$. It  is a Binomial variable with parameter $N(I_\eta(x))$ and $1/2$. The corresponding \pvalue associated to the classical one-sided single test of $\mathcal{H}_0\left\{I_\eta(x)\right\}$ can then be defined by:
\begin{equation}
\label{Eq:pvTwo}
\forall x \in \mathcal{X}_\eta,\,\, p_\eta(x) = F_{\mathcal{B}(N(I_\eta(x)),\frac{1}{2})}(S_\eta(x)).
\end{equation}
This process depends on $N$ and $x$, and once again can be computed very easily.

\subsection{\label{SubSec:Making}Making continuous testing computationally tractable in general}

\begin{figure*}	
	\begin{center}
		\framebox{\includegraphics[scale=0.5]{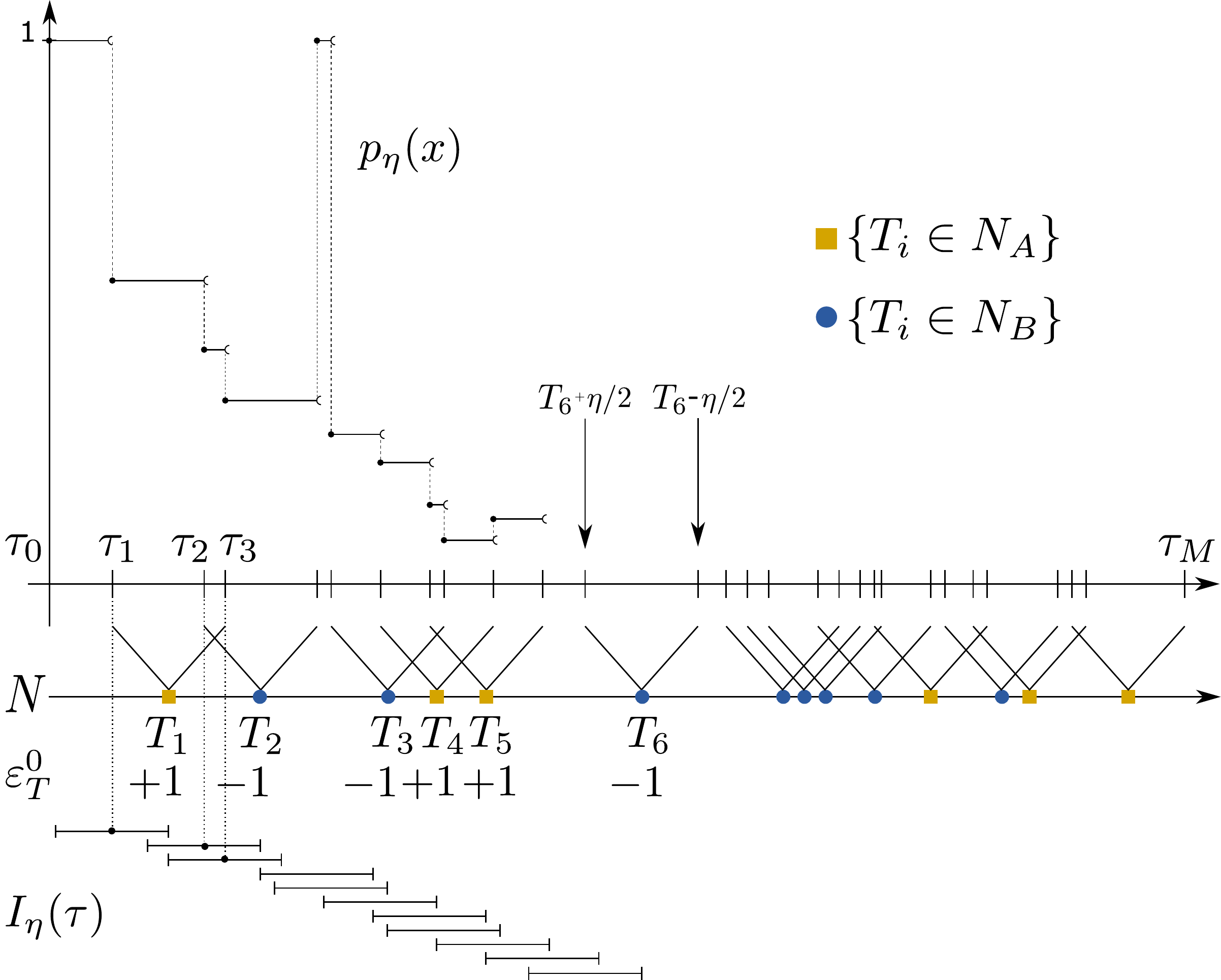}}
	\end{center}
	\caption{Construction of the \pvalue process for the two-sample test. Occurrences from $A$ and $B$ are merged to from the joint process $N$ and Rademacher marks $(\varepsilon_T^0)_T$ are introduced as labels. Each occurrence $T \in N$ has a span $\eta$ that is used to create the partition $\tau$ whose elements constitute the centers of the testing windows $I_\eta(\tau_m)$, $(\tau_m \in \tau)$. Such construction of the windows ensures that the composition of the observed point process is constant between two centers $\tau_m$ and $\tau_{m+1}$ in terms of number and repartition of points. The \pvalue process is then a c\`adl\`ag process with jumps defined by partition $\tau$.\label{Fig:pvalue}
			}
\end{figure*}

Even if the method is based on a {\it continuum} of tests,  the two previous examples are computationally tractable because they are based on $Y(I_\eta(x))$, the number of points within windows $I_\eta(x)$, which can be computed very efficiently and which is piece-wise constant as a function of $x$. In full generality, the number of points may not be informative enough, and to gain power, one can consider more general single test statistics (see Section \ref{kernel}) that only depend on  the \textit{composition} of each window, that is on the function
$$
x\in\mathcal{X}_\eta \mapsto Y\cap I_\eta(x),
$$
which records the random positions of points in $Y$ that lie in $I_\eta(x)$. Since the composition is piece-wise constant with changes only each time a point leaves or enters the scanning window, the single test statistic that only depends on the composition is a piece-wise constant function on a random finite partition $\tau=(\tau_m)_{0 \leq m\leq M}$. Therefore, the \textit{continuum} of tests can be reduced to a random finite number of tests, one for each segment defined by $\tau$. An example of such a partition if provided in Fig. \ref{Fig:pvalue}.

\subsection{General definition of the \pvalue process \label{gendef}}

In the sequel, we assume that a test statistic $S_\eta(x)$ is given for each window $I_\eta(x)$, that only depends on the composition of the window. Because Poisson processes have nice properties of independence between disjoint sets, it is easy to show the following property, which is usually referred to as subset pivotality \citep{WY1993,RW2005}.
\begin{equation*}
\label{equ-P}
 \textrm{\begin{tabular}{c|l} $(\mathcal{P})$ &  For all $\theta$, the conditional distribution of $(S_\eta(x))_{x\in J_0^\eta}$ given $\Nsf$ \\
& is the same whether $Y\sim P_{\theta}$ or  $Y\sim P_{\theta_0}$.
\end{tabular}}
\end{equation*}

This roughly says the the conditional distribution of $S_\eta(x)$ given $\Nsf$ is not modified by the truth or the falsehood of the remaining hypotheses. In particular, the conditional distribution of  $S_\eta(x)$ given $\Nsf$ under $\mathcal{H}_0\left\{I_\eta(x)\right\}$ is the same as the one under the full null $\mathcal{H}_0\left\{[0,1]\right\}$.

Note that in full generality the conditional distribution of $S_\eta(x)$ given $\Nsf$ depends on $\Nsf$ but also on the composition of the window centered on $x$. Therefore we define:
$$\forall x \in J_0^\eta,\,\, F_{\Nsf, x}(s)=P_{\theta_0}(S_\eta(x)\geq s | \Nsf)=P_{\theta}(S_\eta(x)\geq s | \Nsf),$$
and  the associated \pvalue process by
\begin{equation}
  \label{theopvalues}
p_\eta(x) = F_{\Nsf,x}(S_\eta(x)), \,\,   x \in \mathcal{X}_\eta. 
\end{equation}
It satisfies for all $x\in \mathcal{X}_\eta$, for all $(\theta,\lab)$,
\begin{equation} \label{equdefpvalues}
\mbox{if $\mathcal{H}_0\left\{I_\eta(x)\right\}$ is true, }\:\:\forall \alpha\in[0,1], 
\P_{\theta,\lab}({p}_\eta(x)\leq \alpha)\leq \alpha.
\end{equation}

Moreover since $p_\eta(x)$ is a deterministic function of $S_\eta(x)$ (conditionally to $\Nsf$), property $(\mathcal{P})$ also holds for $p_\eta(x)$, and not only for $S_\eta(x)$.

\section{\label{control}False positive control in continuous time}

Except when specifically mentioned, the results presented here hold for any test statistic $S_\eta(x)$ that only depends on the composition of the window (as defined in Section \ref{SubSec:Making}). Their associated \pvalue process are as described previously. 

A rejection set $\mathcal{R}_\eta$ is a subset of $\mathcal{X}_\eta$, which is a function of the \pvalue process $(p_\eta(x))_{x\in\mathcal{X}_\eta}$. Classically, it corresponds to small values of this process and a typical example is given by
\begin{equation*}\label{thresh}
\mathcal{R}_{\eta}(u):=  \left\{ x \in \mathcal{X}_\eta \::\: p_{\eta}(x) \leq u \right\},
\end{equation*}
for a given threshold  $u$ potentially depending on the data. 
Once the rejection set $\mathcal{R}_\eta$ given, the set of accepted windows is defined by its complement
$\mathcal{A}_{\eta}:= (\mathcal{R}_\eta)^c.$

\subsection{Definition of the multiplicity error rates}

If the procedure behaves properly, the set $\mathcal{A}_{\eta}$ should be a good approximation of $J_0^\eta$, or similarly $\mathcal{R}_\eta$ should be a good approximation of $(J_0^\eta)^c$. To evaluate the quality of a procedure $\mathcal{R}_{\eta}$, we focus on \textit{false positive windows} that correspond to elements of $J_0^\eta \cap \mathcal{R}_{\eta}$. The size of $J_0^\eta \cap \mathcal{R}_{\eta}$ can be gauged in many ways. In this work, we focus on the most famous criteria, namely the Family-Wise Error Rate (FWER)  and the False Discovery Rate (FDR).

From a family-wise point of view,  one wants to avoid any false positive with a large probability, which corresponds to a control of the quantity 
\begin{equation} \label{FWER}
\FWER_{\theta,\lab}^\eta(\mathcal{R}_{\eta})=\P_{\theta,\lab}\left( J_0^\eta \cap \mathcal{R}_{\eta}  \neq \emptyset \right).
\end{equation}

While a FWER control provides a strong assessment on the amount of false positives, this approach generally leads to conservative procedures. A more balanced view is to allow some false positives, in a pre-specified fraction of the  total amount of positives. This can be achieved by controlling 
  \begin{equation} \label{FDR}
\FDR_{\theta,\lab}^{\eta}( \mathcal{R}_{\eta})=\E_{\theta,\lab}\left( \frac{\Lambda\left(   J_0^\eta \cap \mathcal{R}_{\eta} \right)}{\Lambda\left(  \mathcal{R}_{\eta} \right)} \right), 
\end{equation}
with $0/0=0$. Importantly, since the setting is continuous here, the ``number" of (false) positives is quantified above by the Lebesgue measure, $\Lambda$, on $[0,1]$.

\subsection{A \minp procedure to control the FWER in continuous time}\label{sec:FWER}

We would like to find a threshold $u^\alpha \in [0,1]$ such that: 
\begin{equation*}
\forall (\theta,\lab), \:\:\FWER_{\theta,\lab}^\eta(  \mathcal{R}_{\eta}(u^\alpha))\leq \alpha. \label{FWERcontrol}
\end{equation*}
For all $u \in [0,1]$,
$$ 
  \left\{J_0^\eta \cap \mathcal{R}_{\eta}(u)  \neq \emptyset \right\} = \left\{\exists x \in J_0^\eta  \::\: {p}_{\eta}(x) \leq   u\right\} \\
=\left\{ \underset{x \in J_0^\eta}{\inf} \left\{ p_{\eta}(x) \right\} \leq u \right\}, $$
with $p_\eta(x)$ defined in \eqref{theopvalues}. 
The set $J_0^\eta$ being unknown, one can use some rough upper-bound. 
We have, thanks to subset pivotality $\left(\mathcal{P}\right)$, that for any $u>0$ (possibly depending on the conditioning variable $\Nsf$),
\begin{eqnarray*}
\FWER_{\theta,\lab}^\eta(\mathcal{R}_{\eta}(u)) &=&\E_{\lab}\left[ \P_{\theta} \left(\underset{x \in J_0^\eta}{\inf} \left\{ p_{\eta}(x) \right\} \leq u\:\big|\: \Nsf\right)\right] \\
&=&\E_{\lab} \left[\P_{\theta_0} \left(\underset{x \in J_0^\eta}{\inf} \left\{ p_{\eta}(x) \right\}  \leq u\:\big|\: \Nsf\right)\right].
\end{eqnarray*}   
Since $J_0^\eta\subset \mathcal{X}_\eta$, we obtain 
$$
  \FWER_{\theta,\lab}^\eta(\mathcal{R}_{\eta}(u)) \leq E_{\lab}\left[P_{\theta_0} \left(\underset{x \in \mathcal{X}_\eta}{\inf} \left\{ p_{\eta}(x) \right\}  \leq u\:\big|\: \Nsf\right)\right].$$
The distribution of $\underset{x \in \mathcal{X}_\eta}{\inf} \left\{ p_{\eta}(x) \right\}$  under $\theta_0$ and given $\Nsf$   are in both cases (homogeneity and two sample cases) known/observable once the variable $\Nsf$ is fixed and it is possible to calibrate an adequate threshold $u^\alpha$. However, due to the classical subtleties of inversion of non continuous c.d.f., it is in fact more powerful to use {\it adjusted $p$-values} that are defined by
\begin{equation}
\label{qvalues}
{q}_{\eta}(x) = F^{\m}_{\Nsf}\left( p_{\eta}(x) \right), \:\:\: x \in \mathcal{X}_\eta,
\end{equation}
with $F^{\m}_{\Nsf}$ being the conditional c.d.f. of $\underset{x \in \mathcal{X}_\eta}{\inf} \left\{ p_{\eta}(x) \right\}$  under the full null hypothesis, that is $\theta=̀\theta_0$.

\begin{theorem}\label{th:FWERSD}
	Let $\alpha \in (0,1)$ and   $q_\eta$ given by \eqref{qvalues}. Then the procedure defined by 
\begin{equation}\label{FWERprocedureq}
	\mathcal{R}_{\eta}^{\m}:= \{x\in \mathcal{X}_\eta\::\: q_\eta(x)\leq\alpha\}
	\end{equation}
has a $\FWER$  controlled by $\alpha$.
\end{theorem}
Note that $\mathcal{R}_{\eta}^{\m}$, which is a continuous version of the classical \minp procedure,  is coherent with the case where there is only one value in the \pvalue process because the resolution $\eta$ is equal to 1. In this extreme case, $\mathcal{X}_\eta$ is reduced to one point and the \pvalue is exactly the adjusted \pvalue, the control of $\FWER(\mathcal{R}_{\eta}^{\m})$ being then an obvious consequence of Section \eqref{equdefpvalues}.

\begin{remark} The previous procedure is usually referred to as a {\it single-step} procedure. We also propose refinements by using a step-down procedure in the Supplementary Material.
\end{remark}

\begin{remark} Unfortunately, even for the classical homogeneity or two-sample tests the c.d.f. $ F^{\m}_{\Nsf}$ is not explicit and needs to be estimated. This is the purpose of Section \ref{boot} in which we also derive meaningful theoretical results when this distribution is approximated by Monte-Carlo that are necessary to provide a completely operational method with strong theoretical guarantee. 
\end{remark} 

\subsection{Link with the scan statistic framework.} 

In the case of the homogeneity test, the \pvalue process does not depend on $x$ and because function $F_{\mathcal{B}(N([0,1]),\frac{1}{2})}$ is a non increasing function, we have 
\begin{equation}
\label{trickscan}
\inf_{x\in\mathcal{X}_\eta} p_\eta(x)= F_{\mathcal{B}(N([0,1]),\frac{1}{2})}\left(\sup_{x\in\mathcal{X}_\eta} S_\eta(x)\right).
\end{equation}
Hence, rejecting for small values of the infinimum of the \pvalue process (similar to standard minP procedures), or rejecting for large of the supremum of the test statistic (i.e. the scan statistic, often called the maxT procedure) is strictly equivalent \citep{DSP03,DV08}. Compared with the work of \cite{CZ07}, we mainly added two ingredients: ($i$) the distribution we are considering is conditional and exact, whereas theirs is asymptotic and still depends on the unknown nuisance parameter $\lab$ (even if they provide a plug-in procedure  in practice to avoid this issue), ($ii$) thanks to Property $(\mathcal{P})$, we are able to prove the FWER control, whereas their control stands under the full null only.

This minP-maxT equivalence is valid in the case of the homogeneity test only, but it does not hold as soon as the c.d.f. of the infinimum of the \pvalue under the full null depends on the composition of the window. This is typically the case for the two-sample test, where $F_{\mathcal{B}(N(I_\eta(x)),\frac{1}{2})}$ depends on $x$ so that Eq. \eqref{trickscan} cannot hold anymore. In this case, the minP framework offers the advantage to scale all tests on the same level \citep{DSP03,DV08}.

\subsection{A weighted BH procedure to control the FDR in continuous time}\label{sec:FDR}

To increase the detection capability, we  propose a second procedure based on the control of the FDR criterion \eqref{FDR}. 
\paragraph{Heuristic.} We would like to find a threshold $v^\alpha \in [0,1]$ (see \eqref{thresh}) such that: 
\begin{equation*}
\forall (\theta, \lab), \FDR_{\theta,\lab}^\eta(v^\alpha)\leq \alpha.\label{FDRcontrol}
\end{equation*}
Following \cite{BDR2014}, let us first explain heuristically how one can reach this goal in a continuous framework.
By Fubini's theorem, $\forall v \in [0,1]$,
$$
\FDR_{\theta,\lab}^{\eta}(v )=\E_{\theta,\lab}\left( \frac{\Lambda\left(   J_0^\eta \cap \mathcal{R}_{\eta}(v) \right)}{\Lambda\left(  \mathcal{R}_{\eta}(v) \right)} \right)\nonumber\\
=\int_{J_0^\eta} \E_{\theta,\lab}\left( \frac{   \ind{ p_\eta(x) \leq v} }{\Lambda\left(  \mathcal{R}_{\eta}(v) \right)}  \right) d\Lambda(x).
$$
Then, choosing $v$ such that $$\frac{ \Lambda\left(  \mathcal{R}_{\eta}(v) \right)}{ \Lambda\left(  \mtc{X}_\eta \right)} \geq \frac{v}{\alpha}$$
and doing as if $v$ was deterministic  leads to
 \begin{equation*}\label{equFDRmaj}
\FDR_{\theta,\lab}^{\eta}(v)\leq \alpha \int_{J_0^\eta} v^{-1}\E_{\theta,\lab}\left( \frac{   \ind{ p_\eta(x) \leq v} }{   \Lambda\left(  \mtc{X}_\eta \right)}  \right) d\Lambda(x),
\end{equation*}
which leads to
\begin{align*}
\FDR_{\theta,\lab}^{\eta}(v)&\leq \frac{\alpha}{\Lambda\left(  \mtc{X}_\eta \right)} \int_{J_0^\eta} v^{-1} \P_{\theta,\lab}\left( p_\eta(x) \leq v \right) d\Lambda(x) \leq \alpha\:\Lambda(J_0^\eta)/\Lambda\left(  \mtc{X}_\eta \right)\leq \alpha,
\end{align*}
by using the  property of \pvalues \eqref{equdefpvalues}. The above heuristic suggests to choose $V^\alpha$ by using 
\begin{equation*}\label{SU}
V^\alpha = \max\left\{ v\geq 0 \::\: \frac{ \Lambda\left(  \mathcal{R}_{\eta}(v) \right)}{ \Lambda\left(  \mtc{X}_\eta \right)} \geq v/\alpha\right\}.
\end{equation*}

\paragraph{Weighted-BH procedure.} Quantile $V ^\alpha$ is random (even conditionally on $\Nsf$) because it depends on the observed \pvalue process. Nevertheless, since the $p$-value process is piece-wise constant  in our framework, finding threshold $V^\alpha$ is very simple. We propose a step-up algorithm that is inspired from the famous {\it weighted BH threshold} \citep{BH1997}, with the essential difference that our algorithm is based on random weights. Recall that the processes $(S_\eta(x))_x$ and therefore $(p_\eta(x))_x$ are piece-wise constant on the random partition $\tau=(\tau_m)_{0 \leq m\leq M}$ ($\tau_0=0$, $\tau_M=1$). Therefore
$$
  \Lambda\left(  \mathcal{R}_{\eta}(v) \right)
  =\sum_{m=1}^{M} (\tau_{m}-\tau_{m-1}) \ind{p_\eta(\tau_{m-1})\leq v}.
$$
This implies that the threshold $V^\alpha$ defined  in \eqref{SU} can be derived as follows:
\begin{itemize}
\item Compute the weights, 
$$w_m=\frac{\tau_{m}-\tau_{m-1}}{\Lambda\left(  \mtc{X}_\eta \right)},\
,\, 1\leq m \leq M, $$

\item Denote 
$p_m=p_\eta(\tau_{m-1})$, $1\leq m \leq M$,

and order these \pvalues  in increasing order $p_{\sigma(1)}\leq \dots \leq p_{\sigma(M)}$ for an appropriate permutation $\sigma$ of $\{1,\dots,M\}$;
\item Consider $\widehat{k}=\max\{k\in\{1,\dots,M\}\::\: p_{\sigma(k)}\leq \alpha \sum_{\l=1}^k w_{\sigma(\l)} \}$ (with $\widehat{k}=0$ if the set is empty);
\item Compute $V^\alpha$ as $\alpha \sum_{\l=1}^{\widehat{k}} w_{\sigma(\l)}.$
\item Define $\mathcal{R}_{\eta}^{wBH}=\{x \in \mtc{X}_\eta : p_\eta(x)\leq V^\alpha\}.$
\end{itemize}

By contrast with the classical BH step-up procedure, each $p$-value has its own weight $w_m$ which is not uniform but  adaptive to the data. This weight means that  the quantity of interest is not the number of tests, i.e. the cardinality of the partition $\tau$, but the proportion of "time" that the \pvalue process spends at a particular value.

\begin{remark}It is worth to note that  $\mathcal{R}_{\eta}^{wBH}=\{x \in \mtc{X}_\eta : q_\eta(x)\leq \alpha\}$ 
where ${q}_\eta(x)$ stands for the adjusted $p$-value
 $${q}_\eta(x) = \min_{k: p_{\sigma(k)}\geq p_\eta(x)} \left\{ \frac{p_{\sigma(k)}}{ \sum_{\l=1}^{{k}} w_{\sigma(\l)}} \right\}.$$
 \end{remark}

\paragraph{Theoretical result.} Our procedure is based on a heuristic that has been made fully theoretically valid by \cite{BDR2014} under suitable assumptions of measurability and  positive dependence (namely, finite dimensional positively regressively dependent $p$-value process). The difficulty comes from the interplay between the numerator and the denominator within the expectation in \eqref{equFDRmaj} via $v$, because $v=V^\alpha$ is a random variable. While the measurability issue is not critical here (by the piece-wise constant property), the positive dependence condition  seems difficult to check in general. For instance, this condition is known to exclude the two-sided test statistic \citep{Yek2008}. We prove here that this positive dependence condition holds in the one-sided testing setting (both for the homogeneity test with a known $\lambda$ and the two-sample test), by using the peculiar properties of Poisson processes.

\begin{theorem}\label{th-FDR}
	For all $\alpha \in (0,1)$, for a \pvalue process given either by Eq. \ref{Eq:pvHom} or Eq. \ref{Eq:pvTwo}, for one-sided null hypotheses, the corresponding procedure $\mathcal{R}_\eta^{wBH}$, satisfies
	$$\forall (\theta,\lab), \:\:\FDR_{\theta,\lab}^\eta(\mathcal{R}_\eta^{wBH})\leq \alpha.$$
\end{theorem}

This result is proved in the Supplementary File. Theorem~\ref{th-FDR} hence theoretically justifies the use of a weighted step-up procedure to control the continuous FDR. However, it is only valid for simple versions of our test statistic, which ensure the positive dependence structure. For more general cases, \cite{BDR2014} proposed a modification of the step-up algorithm. This modification consists in introducing a function $\beta(v)\leq v$ of a specific form, such that the quantity $\sum_{\l=1}^k w_{\sigma(\l)}$ is replaced by the smaller quantity $\beta\left(\sum_{\l=1}^k w_{\sigma(\l)}\right)$. However, this modification is reported to be quite conservative \citep{BR2008}. Furthermore, our simulation results show that the FDR control \eqref{FDRcontrol} is achieved in most of the situations we studied, even for more complex \pvalue processes that the ones we considered here (homogeneity test with known intensity, and two-sample test)

Note that the homogeneity test with known intensity in Theorem~\ref{th-FDR} is similar to Corollary 4.5 in \cite{BDR2014}. However, a subtle distinction concerns  the way the null hypotheses are defined (windows versus points). A consequence is that  our approach does not rely on any knowledge regarding the regularity of the underlying process intensity.

\section{\label{kernel} Kernel-based statistics for sliding windows}

\subsection{On the limitations of the count-based statistic}

Let us focus on the two-sample test to illustrate our motivations. In this context, our objective is to detect local differences between unknown functions $\intA$ and $\intB$ based on individual tests defined on a set of windows. In a first approximation, this can be achieved by comparing $N_A(I)$ with $N_B(I)$, the count of occurrences on window $I$, that are distributed as a couple of independent Poisson variables. However, given that $N_A(I)$ (resp. $N_B(I)$) is an approximation of $\int_I \intA(t) dt$ (resp. $\int_I \intB(t) dt$), the count based-statistic strategy can be considered as an hypothesis that tests the equality of function integrals rather than the equality of the functions themselves. Hence the induced testing procedure may lack of accuracy.

Therefore we developed a kernel-based strategy to propose single test statistics that do not depend on the sole number of occurrences in a window, but that consider its whole composition as defined in Section \ref{SubSec:Making}. The interest of kernel-based statistics is that they account the physical distances between occurrences, which explains the gain of performance.

\subsection{The kernel vision in the two-sample case}

The kernel framework for testing proposed by \cite{GBR12} and \cite{FLRB13} has been especially designed to separate functions, and not just their integrals. It is based on the estimation of a distance between the functions $\intA$ and $\intB$ (in a RKHS, or in $L^2$). In particular, \cite{FLRB13} show that these test statistics are minimax with respect to various smoothness classes, and in this sense, powerful against some alternatives that cannot be distinguished by count-based tests. 

In the following we consider translation invariant kernels denoted by $K$, and in practice we will restrict ourselves (see Sections \ref{sim} and \ref{appli}) to the Gaussian kernel that has been shown to be the most powerful on simulations \citep{FLRB13}. The shape of the statistic is then given by $$S_\eta(x)=\sum_{T\neq T' \in N\cap I_\eta(x)} K(T-T') \eps_T\eps_T',$$
where $N$ is the joint process and  the $\eps_T$'s are the marks (see Section \ref{frame}).  In particular, it only depends on the composition of the window. To understand, why this statistic is meaningful, let us compute its expectation. By classical computations for Poisson processes, 
\begin{eqnarray*}
E_{\theta,\lab}(S_\eta(x))&=&E_\lab \left[\sum_{T\neq T' \in N\cap I_\eta(x)} K(T-T') \theta(T)\theta(T') \right]\\
&=& \int_{I_\eta(x)^2} K(t-t') \theta(t)\theta(t') \lab(t) \lab(t') dt dt'\\
&=&  \int_{I_\eta(x)^2} K(t-t') [\intA(t)-\intB(t)][\intA(t')-\intB(t')] dt dt'
\end{eqnarray*}
If $K\star f$ denote the convolution product, we end up with
$$E_{\theta,\lab}(S_\eta(x)) = \int_{I_\eta(x)} \left(K\star\left[(\intA-\intB){\bf 1}_{I_\eta(x)}\right]\right)(t) [\intA(t)-\intB(t)] dt.$$
In particular if $K$ is a convolution kernel with bandwidth $h$ we obtain, under classical assumptions, that 
$$
E_{\theta,\lab}(S_\eta(x)) \underset{h \rightarrow 0}{\longrightarrow} \int_{I_\eta(x)} [\intA(t)-\intB(t)]^2 dt.
$$
This short derivation illustrates the motivation to use such a kernel-based statistic, as a good estimate of a distance between intensity functions. The one-sided version of this statistic, as well as the homogeneity case are presented in the Supplementary Material. Note that from a computational point of view, the complexity for computing $S_\eta(x)$ is in $O(N^2(I_\eta(x))$. However, thanks to the piece-wise constant property of $S_\eta(x)$ on partition $\tau$, the continuum $\{S_\eta(x),x\in \mathcal{X}_\eta\}$, can be computed using the in/out points of each scanning window, which drastically reduces the global computational burden.

\section{\label{boot} Monte-carlo procedures for \pvalues estimation and adjustment}
\begin{table*}
\begin{tabular}{|c|c|c|cc|}
\hline
Procedure			   &  test statistic             &  computations    & homogeneity test & two-sample test\\
\hline				   
\multirow{4}{*}{\minp} & \multirow{2}{*}{count} & $p_\eta$       & \multicolumn{2}{c|}{exact}     \\
					   & 							 & full procedure & \multicolumn{2}{c|}{1-step MC} \\
					   & \multirow{2}{*}{kernel} & $p_\eta$      & \multicolumn{2}{c|}{1-step MC} \\
					   & 							 & full procedure & \multicolumn{2}{c|}{2-step MC} \\
\hline					   
\hline
\multirow{4}{*}{wBH}   & \multirow{2}{*}{count} & $p_\eta$       & \multicolumn{2}{c|}{exact}     \\
					   & 							 & full procedure & \multicolumn{2}{c|}{exact} \\
					   & \multirow{2}{*}{kernel} & $p_\eta$      & \multicolumn{2}{c|}{1-step MC} \\
					   & 							 & full procedure & \multicolumn{2}{c|}{1-step MC} \\					   					   
\hline				   
\end{tabular}
\caption{Summary of the required computations to implement the continuous tests in practice for the homogeneity and two-sample test.\label{Tab:procMC}}
\end{table*}

\subsection{The distribution of the single test statistic being explicit.} 

For tests based on the count statistic the \pvalue process is explicitly computable. Hence, no need for Monte-Carlo in this case. Moreover, since our weighted BH procedure (Section \ref{sec:FWER}) is completely explicit, $\mathcal{R}^{wBH}_\eta$ can be easily computed in practice. In this situation we have already proved the FDR control for some particular cases and we will show that it holds in more general cases by using simulations (Section \ref{sim}). 

However, the \minp procedure is not explicit since it requires the conditional distribution of $\inf_{x \in \mathcal{X}_\eta} p_\eta(x)$ given $\Nsf$ under the full null hypothesis (Section \ref{sec:FWER}). Therefore we need to use a Monte-Carlo procedure to approach this distribution. This can be achieved while maintaining a valid FWER control by following \cite{RW2005}, as detailed in Section \ref{sec:twostepMC} below.

\subsection{The distribution of the single test statistic being not explicit} 

This concerns kernel-based tests (Section \ref{kernel}) for which the \pvalue process is not even computable. One can use again the estimated \pvalues developed by \cite{RomanoWolf} since their result shows that they satisfy \eqref{equdefpvalues} and $\mathcal{R}^{wBH}_\eta$ can be easily computed in practice. However, there is no theoretical FDR guarantee and we will only show the FDR control by simulations in Section \ref{sim}. Consequently, to compute $\widehat{p}_\eta(x)$, the weighted-BH procedure only requires one Monte-Carlo step in this situation.

As for the \minp procedure (Section \ref{sec:FWER}), it requires an additional Monte-Carlo step. To this end, a classical and sufficient method is to perform an approximation using two independent Monte-Carlo samples. We detail this approach for the Homogeneity case in the supplementary material. For the two-sample case, the Monte-Carlo procedure is quite demanding and is close in spirit to bootstrap methods. In this case, we developed a double Monte-Carlo procedure, which only requires \textit{one} Monte-Carlo sample and still guarantees a controlled FWER (Section \ref{sec:twostepMC}).

\subsection{Estimation of the \pvalue process by Monte-Carlo} 

Recall that this estimation is required when using test statistics different from the statistics based on counts. The full procedures are summarized in Table \ref{Tab:procMC}.

\paragraph{Homogeneity test.} Here, the conditioning variable $\Nsf=N([0,1])$ is the same for all scanning windows and the distribution of $S_\eta(x)$ under $\mathcal{H}_0\{I_\eta(x)\}$ does not depend on $x$. Let us focus on the first window $I_\eta(\eta/2)=(0,\eta]$ and simulate points inside this window under $\mathcal{H}_0\{I_\eta(\eta/2)\}$.
Hence we consider $B$ random samples of the number of points $N^b(I_\eta(\eta/2))$ such that 
$$
\forall b \in \{1,\hdots,B\},\,\, N^b(I_\eta(\eta/2)) | \big\{N([0,1])=n\big \} \underset{iid}{\sim} \mathcal{B}\big(n,\eta\big),
$$
and then we draw $$\big(T_1,\hdots,T_{N^b(I_\eta(\eta/2))}\big) \underset{iid}{\sim} U((0,\eta]),$$
to sample positions of a homogeneous Poisson process $N^b$ inside window $(0,\eta]$. For each $b$, one can compute thanks to the positions of the points inside $[0,\eta]$ the value of the test statistic for the first window, that we denote $S_\eta^b$. But since all the statistics are distribution invariant by translation under the null, one can also say that this sample has the same distribution as $S_\eta(x)$ under $\mathcal{H}_0\{I_\eta(x)\}$ for every window center $x\in\mathcal{X}_\eta$. 

Therefore, whatever $B$, one has easily access to $\big(S^1_\eta \hdots, S^B_\eta\big)$, $B$ i.i.d. variables with the same distribution as $S_\eta(x)$ under $\mathcal{H}_0\{I_\eta(x)\}$ for every $x$. Then, following \cite{RomanoWolf}, we define:
\begin{equation*}
\label{MCpvaluesHomo}
\widehat{p}_{\eta}(x) = \frac{1}{B+1} \left(1+\sum_{b=1}^{B} \ind{S_\eta^b \geq S^0_\eta(x)} \right),
\end{equation*}
which amounts to use the empirical function $F_{\Nsf,x}$ defined in Eq. \ref{theopvalues} over a $B+1$ sample where $S^0_\eta(x)$ is the observed statistic.

\paragraph{Two sample test.} We label the observed marks $\eps^0:=\eps=(\eps_T)_{T\in N},$ such that it constitutes the first term of a $(B+1)$-sample of marks,  filled by $B$ independent draws of \textit{i.i.d.}  Rademacher sets, $\eps^b:=(\eps^b_T)_{T\in N}$, for $b=1,...,B$. The conditional distribution given $N=\Nsf$ of these Rademacher processes are:
$$
\eps^b_T | N \underset{iid}{\sim} 2\mathcal{B}\left( \frac{1}{2}  \right)-1,
$$
that is the distribution that the observed marks $\eps_T$'s should have under the full null hypothesis. As previously, the Monte-Carlo approximated conditional \pvalue of window $I_\eta(x)$ is defined  by:
\begin{equation}
\label{MCpvalues}
\widehat{p}_{\eta}(x) = \frac{1}{B+1} \left(1+\sum_{b=1}^{B} \ind{S_\eta^b(x) \geq S_\eta^0(x)} \right),
\end{equation}
where $S^0_\eta(x)$ is the observed statistic and where $S_\eta^b(x)$ is the value of the test statistic  on $I_\eta(x)$ that is computed with the resampled marks $\varepsilon^b$ and fixed joint process $N$. Let us underline that because in the two-sample case the partition $\tau$ only depend on $N$ and do not vary with the resampling scheme $\varepsilon^b$, the processes $(S^b_{\eta}(x))_{x\in \mathcal{X}_\eta}$, for $b=0,...,B$, and therefore the process $(\widehat{p}_{\eta}(x))_{x\in \mathcal{X}_\eta}$ are piece-wise constant on $\tau$. 

\subsection{Two-step Monte-Carlo method for the \minp procedure.\label{sec:twostepMC}} 

Recall that when using the \minp procedure with a \pvalue process that is not explicit, we need two Monte-Carlo approximations. 

\paragraph{Homogeneity test.}
In that case, the distribution of the test statistics under the null are independent of the window centers, while the distribution of the minimum of the $p$-values relies on the observations over all windows. Hence, it is quite natural to perform two separated Monte-Carlo schemes for these two steps.

\paragraph{Two sample test.} 

In the two-sample case, the trick is that one can use the \textit{same sample} of $(N,\eps^b)$'s as the one used in \eqref{MCpvalues} for both approximations. More precisely, the \pvalues  $\widehat{p}_\eta^b(x)$, corresponding to the simulated process $(N,\eps^b)$, can be computed as
\begin{equation*}\label{simpval}
\forall b \in \{0,\dots,B\},\, \forall x \in \mathcal{X}_\eta,\,\, \widehat{p}^{b}_{\eta}(x) = \frac{1}{B+1} \sum_{b'=0}^{B} \ind{S_\eta^{b'}(x) \geq S_\eta^b(x)},
\end{equation*}
that is we do the same computation as in Eq. \ref{MCpvalues} except that we replace the observed statistics $S^0_\eta(x)$ by the  simulated one $S^b_\eta(x)$, hereby computing a "resampled" p-value process corresponding to $\varepsilon^b$. Then, let us define, for all $b=0,...,B$
$$m^{b}=\inf_{x \in \mathcal{X}_\eta}\widehat{p}^{b}_{\eta}(x).$$
The corresponding adjusted Monte-Carlo \pvalues are then given by
\begin{equation*}    
 \label{equ:adjpvalues}
    \wh{q}_\eta(x) =   \frac{1}{B+1} \left( 1 + \sum_{b=1}^{B} \ind{ m^b \leq \widehat{p}_{\eta}(x)}\right).
 \end{equation*}
 Finally the rejection set is given by
\begin{equation*}\label{Rhatminp}
\widehat{\mathcal{R}}_\eta^{\m} =\{x \in \mathcal{X}_\eta, \widehat{q}_{\eta}(x) \leq \alpha\}.
\end{equation*}

The previous double Monte-Carlo procedure is based on the use of the same random signs $\left(\eps^b_T\right)_{T \in N}$, $0\leq b\leq B$ and not on two sets of different simulations. Indeed the $\eps^b$'s  are used twice: a first time to estimate the $p$-values of the observed process and a second time  to give the \pvalues even on the simulated realizations leading to the adjusted \pvalues. This procedure has the double advantage of sparing computational resource,  while enjoying the same control property as shown hereafter.

\begin{theorem}
\label{th:pvaltwoMC} 
Let $\alpha\in (0,1)$. For the two-sample case, the continuous testing procedure defined by \eqref{Rhatminp} satisfies
$$\forall (\theta,\lab), \FWER_{\theta,\lab}^\eta(\widehat{\mathcal{R}}_\eta^{\m}) \leq \alpha.$$
\end{theorem}
This result holds whatever the choice of the single test statistic that only depends on the composition of the window.

\section{\label{sim} Simulations}

\subsection{Homogeneity test}
For a fixed parameter $r$, we simulate a homogeneous Poisson process corresponding to the following piecewise constant signal $\theta$ (illustrated in Figure \ref{Fig:theta}, Supp. Material):
$$
\theta(t) = 
\begin{cases}
+\theta^*,\ \forall t \in \mathcal{I}_1^+, 
\\
-\theta^*,\ \forall t \in \mathcal{I}_1^-
,
\\
0,\  \text{otherwise}. \\
\end{cases}
$$
with 
$$
\begin{cases}
\mathcal{I}_1^+ =[\frac{1}{4}-\frac{r}{4},\frac{1}{4}+\frac{r}{4}] \cup
[\frac{1}{2}-\frac{r}{4},\frac{1}{2}+\frac{r}{4}] \cup
[\frac{3}{4}-\frac{r}{4},\frac{3}{4}+\frac{r}{4}], \\
\mathcal{I}_1^-=
[\frac{1}{4}-\frac{r}{2},\frac{1}{4}-\frac{r}{4}] \cup
[\frac{1}{4}+\frac{r}{4},\frac{1}{4}+\frac{r}{2}] \\ 
\hspace{0.6cm}\cup [\frac{1}{2}-\frac{r}{2},\frac{1}{2}-\frac{r}{4}] \cup
[\frac{1}{2}+\frac{r}{4},\frac{1}{2}+\frac{r}{2}] \\ 
\hspace{0.6cm}\cup [\frac{3}{4}-\frac{r}{2},\frac{3}{4}-\frac{r}{4}] \cup [\frac{3}{4}+\frac{r}{4},\frac{3}{4}+\frac{r}{2}].
\end{cases}
$$
and $\mathcal{I}_1=\mathcal{I}_1^+ \cup \mathcal{I}_1^-$, $\mathcal{I}_0=[0,1] \backslash \mathcal{I}_1$, the positions on which $\theta(.)$ is not null (or null, respectively). We consider this particular shape of $\theta$ to ensure that $\int_0^1\theta(t)dt=0$. 

For a fixed background intensity parameter $\nu^*$, we therefore simulate 1000 times a Poisson process $N$ of intensity $\nu(.)= \nu^*(1+\theta(.))$. Parameter $\theta^*$ is used to tune the distance to homogeneity (which reflects the difficulty of the test), with $\theta^*=0.01$ standing for a situation where almost every position is under the homogeneous regime, $\theta^*=0.99$ standing for a situation with high separability. Parameter  $\nu^*$ controls the number of occurrence and varies in $\{500,1000\}$. To compare with other scanning procedures, we opt for one-sided tests (the alternative corresponding to too large $\theta$), with fixed window size $\eta=2r$. More precisely, we compare our count-based \minp and count-based weighted BH procedures in continuous time with asymptotic approaches based on the scan statistic that controls the FWER \citep{CZ07} and the FDR \citep{SZY2011}. 

\begin{figure*}
	\begin{center}
		\begin{tabular}{cc}
		\includegraphics[scale=0.25]{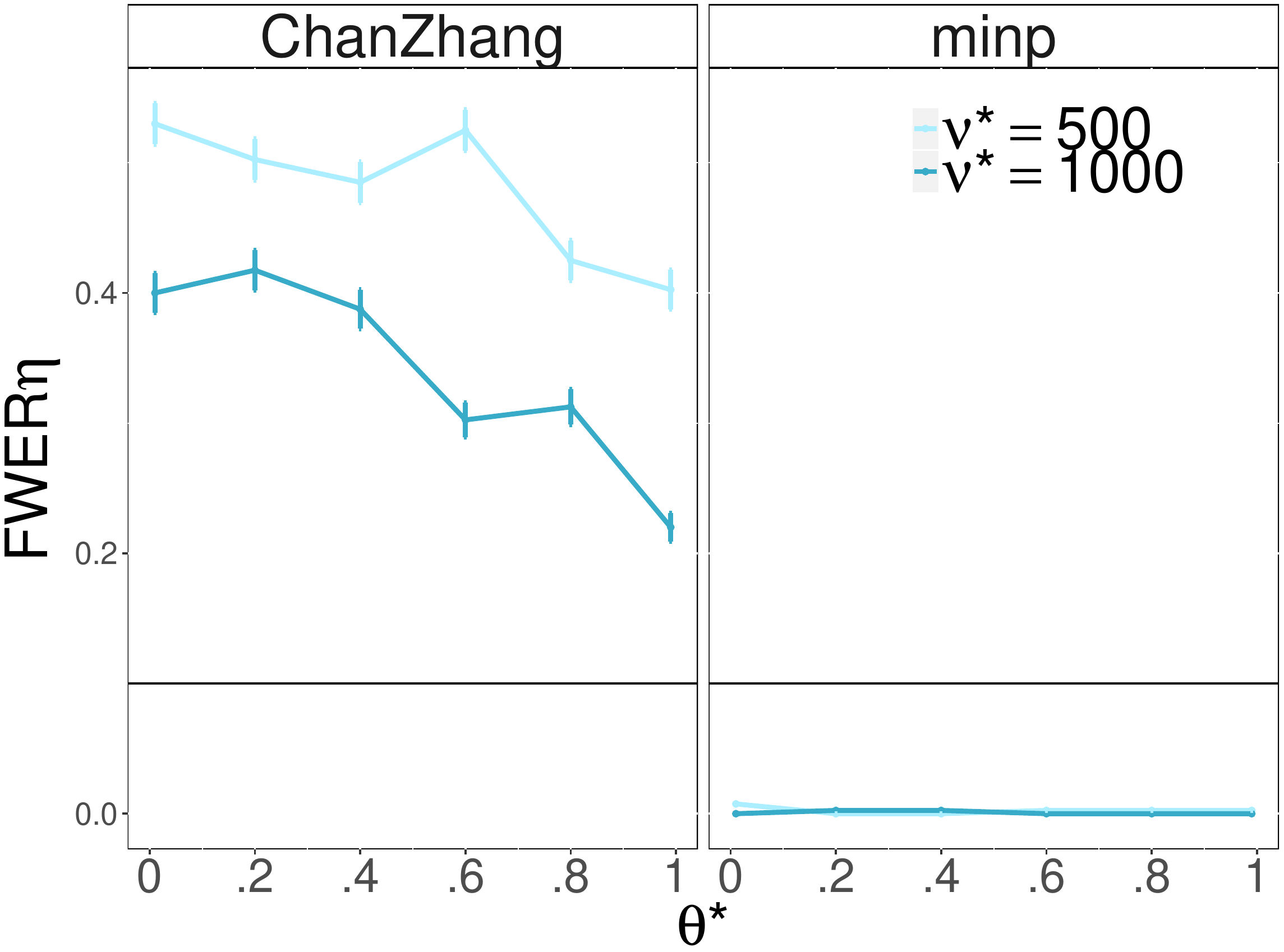} &
		\includegraphics[scale=0.25]{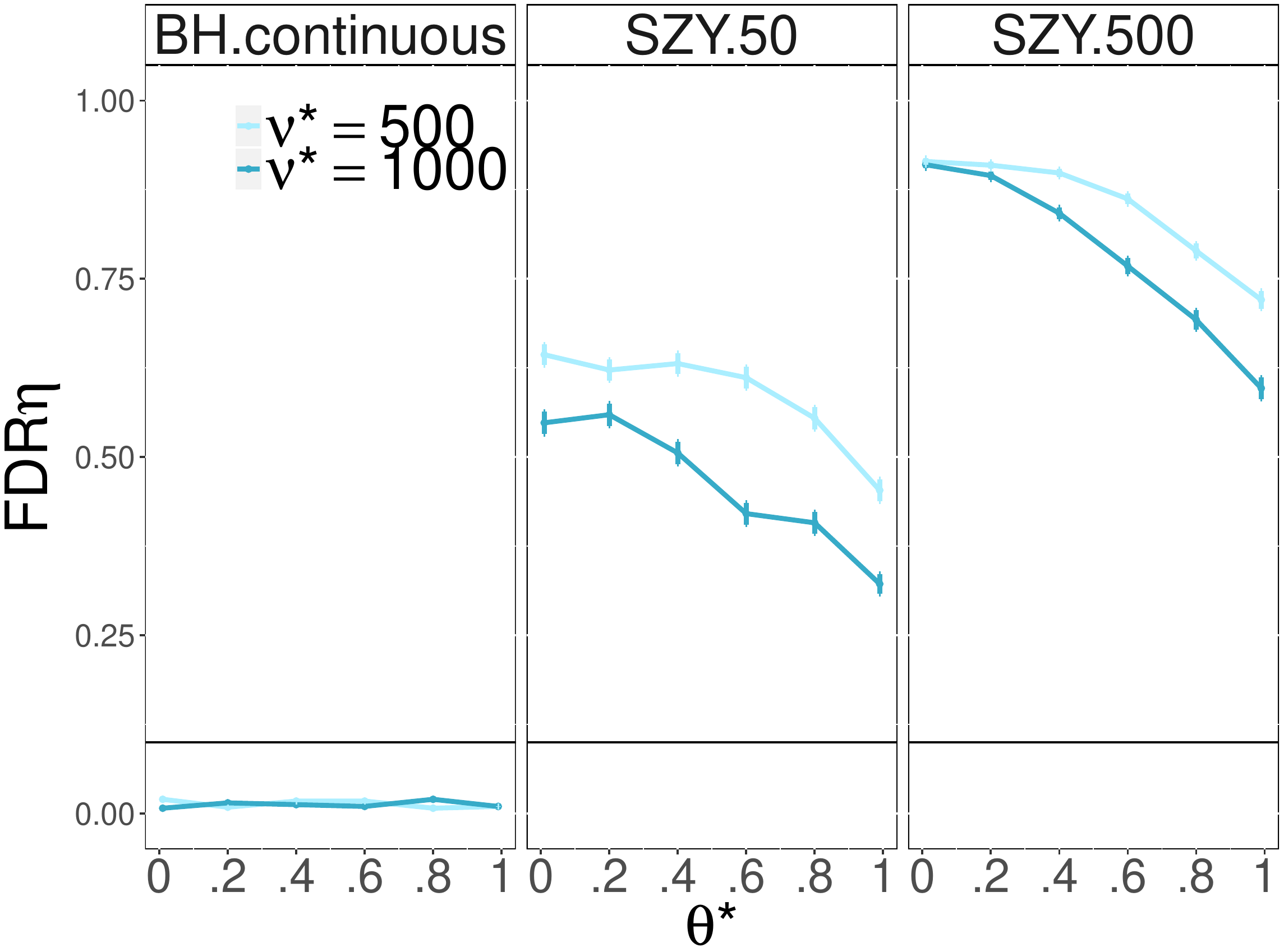}
		\end{tabular}
	\end{center}
	\caption{Comparison of asymptotic vs. non asymptotic methods for the control of Type-I error rates ($\alpha=0.1)$. ChanZhang: asymptotic approximation of the scan statistics distribution proposed by \cite{CZ07}, SYZ: approximation of the FDR proposed by \cite{SZY2011} with 50 (SZY.50) and 500 (SZY.500) windows of identical length. FWER$_\eta$ and FDR$_\eta$ refer to Eqs. \ref{FWER} and \ref{FDR}.\label{Fig:homogeneity_control}}
\end{figure*}

We consider this homogeneity test to compare asymptotic approaches based on the scan statistic to control the FWER \citep{CZ07} and the FDR \citep{SZY2011}, with our \minp and weighted BH procedures in continuous time, based on the count statistic. Figure \ref{Fig:homogeneity_control} clearly shows that our \minp procedure in continuous time controls the FWER even for moderate sample sizes, whereas the scan statistic procedure requires a strong signal ($\theta^* \sim 1)$, or a higher number of occurrences ($\nu^*=5000$, not shown). This trend is even more clear for the FDR control, that is not ensured by method of \cite{SZY2011}. Note that contrary to continuous testing, this method requires a discretization step on which the performance depend. In this case we choose fixed-length non-overlapping windows to compute their procedure. 

\subsection{Two sample test}

To test the differences between intensities $\nu_A$ and $\nu_B$, we sample a joint point process $N$ over $[0,1]$ with constant intensity $\nu(.)=\nu^* \in \{500,1000\}$. Then we sample the labels $\eps^0$ given the occurrences of the joint process $T \in N$, in an independent manner, such that
$$
\forall T \in N, \varepsilon_T^0 | N \sim
\begin{cases}
2\mathcal{B}\big((\theta^*+1)/2\big)-1 & \text{ if } T \in \mathcal{I}_1^, \\
2\mathcal{B}(1/2)-1 & \text{ if } T \in \mathcal{I}_0^.
\end{cases}
$$
Thus positions $T \in N$ with $\eps_T^0=+1$ (resp. $-1$) stand for points of process $N_A$ (resp. $N_B$) with intensity $\nu_A$ (resp. $\nu_B$). Parameter $\theta^*$ represents the distance between $\nu_A$ and $\nu_B$ (which reflects the difficulty of the test), with $(\theta^*+1)/2=0.51$ standing for a situation where the two intensities are nearly equal, and $(\theta^*+1)/2=0.9$ standing for a situation with high separability. $\mathcal{I}_0$ and $\mathcal{I}_1$ are defined as in the homogeneity case, and each configuration is repeated 1000 times. For each value of $\theta^*$, bootstrap \pvalues are computed using  $B=10^5$ Monte Carlo samples, along with the Gaussian kernel $K$ of bandwidth $h=\eta$. Decision rules are based either on the \minp procedure, or on the weighted BH procedure, and we are comparing the performance of the count statistic (one-sided) with the statistic based on the Gaussian Kernel (one-sided). The size of the windows is fixed $\eta=2r$.

\begin{figure*}
	\begin{center}
		\begin{tabular}{cc}
		\includegraphics[scale=0.25]{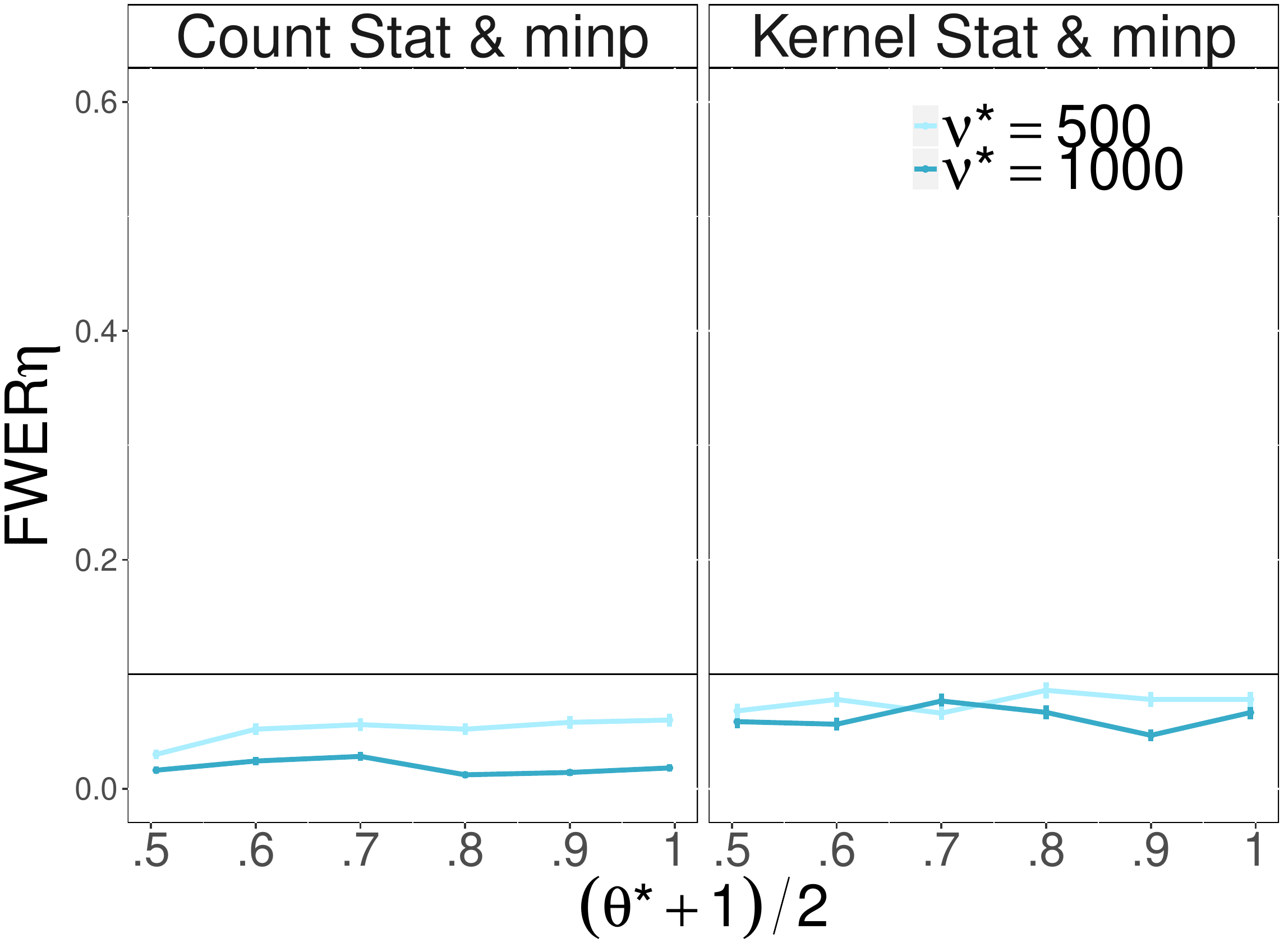} & 
		\includegraphics[scale=0.25]{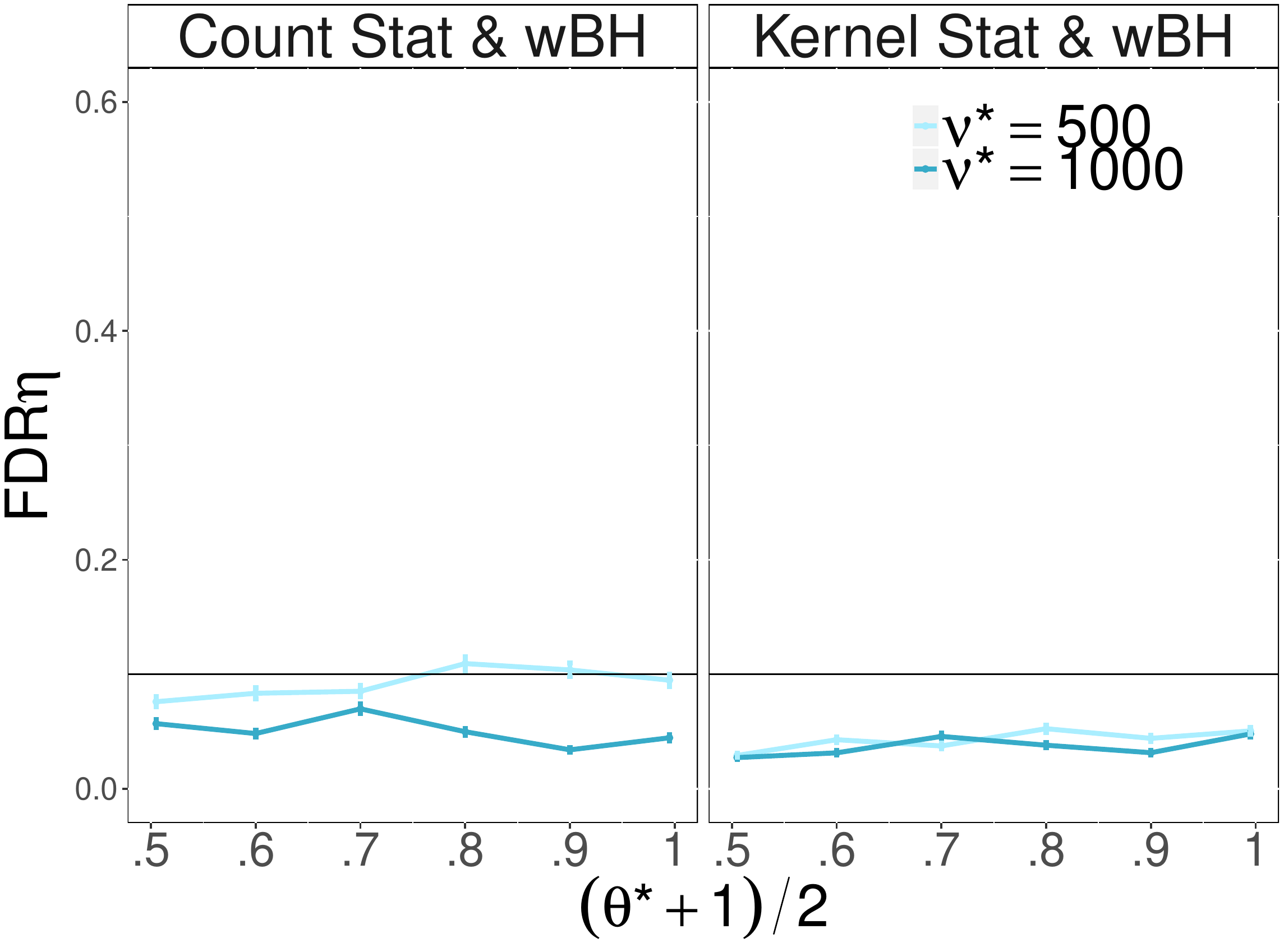} 
	\end{tabular}
	\end{center}
	\caption{Comparison of the count and the Gaussian kernel statistics for the control of joint error rates ($\alpha=0.1$).\label{Fig:SimuKernel}}
\end{figure*}

\begin{figure*}
	\begin{center}
		\includegraphics[scale=0.35]{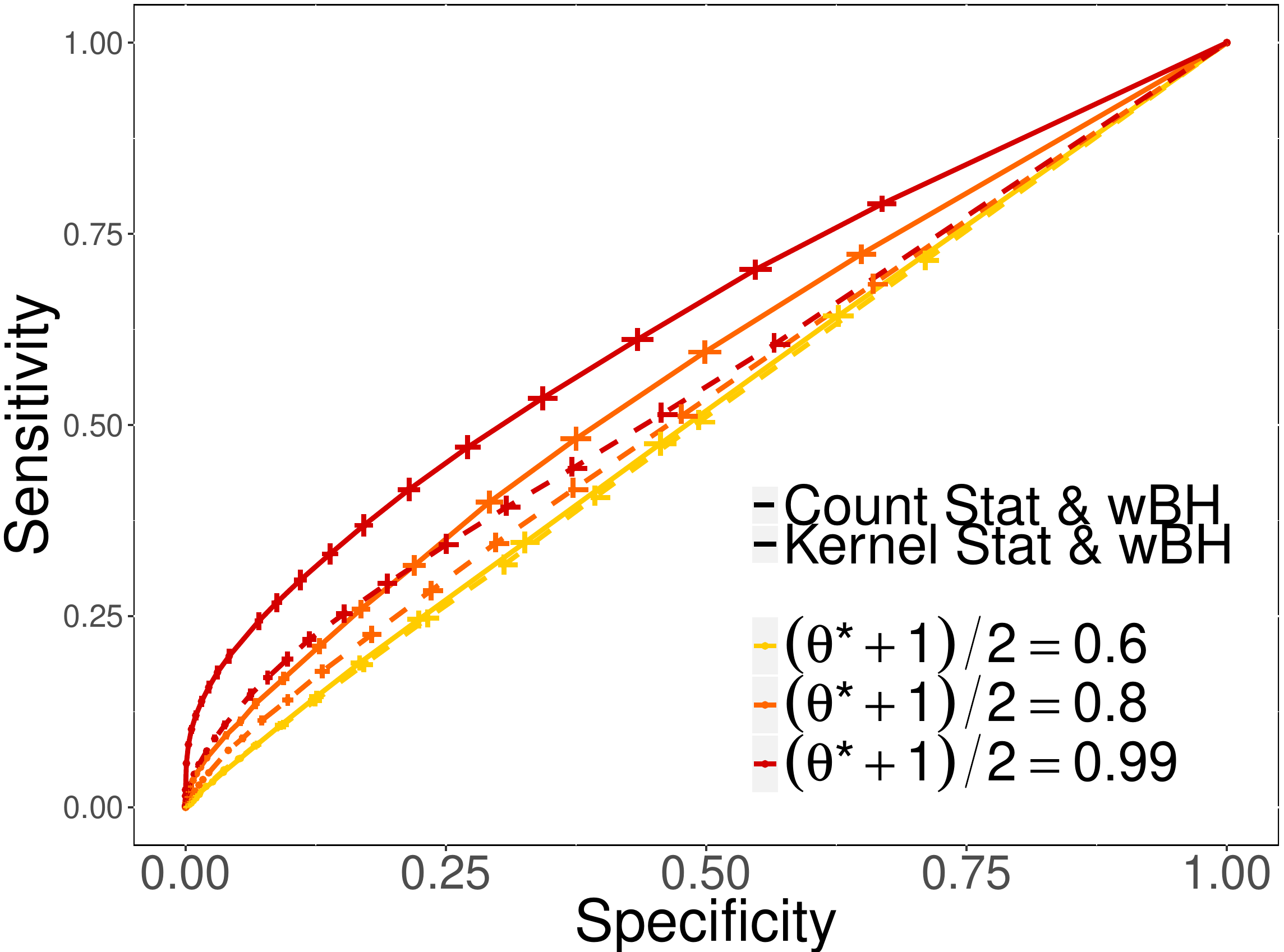}
	\end{center}
	\caption{ROC Curve to compare the Count statistic (dashed lines) with the Gaussian Kernel statistics (plain lines). Sensitivity, or true positive rate (proportion of true positive windows among all rejected windows). Specificity, or true negative rate,  proportion of true negative windows among all accepted windows. One dot corresponds to a value for $\alpha$. Here $\nu^*=1000$, and \pvalues were adjusted using the wBH procedure. \label{Fig:ROC}}
\end{figure*}

Even if the FWER and the FDR are well controlled by our two procedures, the Gaussian kernel statistic seems to be less conservative when controlling the FWER and seems more robust changes in the intensity $\nu^*$ of the process (Fig. \ref{Fig:SimuKernel}). The control of the FDR are comparable between the two statistics. However, when the kernel and the count statistics are compared using sensitivity and specificity (Fig. \ref{Fig:ROC}), for the same level of signal (same $\theta^*$), the kernel statistics is globally more accurate than the count statistics (higher sensitivity and specificity). This can be explained by the fact that kernel-based statistics are in fact $U$-statistics, hereby using information from pairs of points through the kernel. The difference of these statistics will also be illustrated on the analysis of experimental data.

\section{Applications \label{appli}}

\paragraph{Description of the full procedure in practice.} In the following applications, we compute the \pvalue process as well as its \minp and wBH adjusted versions. In order to propose a completely operational procedure, we need to clarify the distinction between the points space and the windows space on which the decision is made. Indeed, if window $I_\eta(x)$ is accepted by our procedure, and if there is no mistake, then one is sure that $H_{0,t}$ holds for all point t in $I_\eta(x)$, whereas if a window $I_\eta(x)$ is rejected without mistake, it is yet not clear whether there is not a smaller interval included in $I_\eta(x)$ on which the null hypothesis holds. Consequently we start by defining the set of accepted windows
$$
\widehat{\mathcal{A}}_\eta = \left\{x \in \mathcal{X}_\eta: \widehat{q}_\eta(x)>\alpha\right\},
$$
with $\widehat{q}_\eta(x)$ denoting the estimated adjusted \pvalue process. Finally, the graphical representation of our procedure is based on the estimated set of points that lie within rejected windows, namely: $\widehat{\mathcal{I}}_1^\eta = [0,1] \backslash \widehat{\mathcal{I}}_0^\eta$,
where 
$$
\widehat{\mathcal{I}}_0^\eta = \left\{t \in [0,1] \::\: \exists x \in \widehat{\mathcal{A}}_\eta \mbox{ s.t. }t\in I_\eta(x) \right\}.
$$

\subsection{\label{real} Do neurons respond to some stimulus ?}

In Neurosciences, the succession of action potentials of a given neuron, also called spike trains, constitutes the key biological signal that carries information regarding the functioning of neurons. In particular, understanding the rapid changes in firing rates due to a certain stimulus \citep{KassVenturaCai} is a challenging task, and inhomogeneous Poisson processes are classically used to model such data \citep{Pipa2013}. Asking whether there is a change of firing rate after exposure typically refers to our homogeneity test, and we are seeking for time intervals on which the rate is modified with respect to the standard behaviour. To illustrate our procedure we analyzed public experimental data \citep{PouzatChaffiol} that consist in  spike trains of a cockroach neuron stimulated by different odors like citronellal or terpineol. A first question is to determine if this neuron responds to the stimulus (one-sided homogeneity test), and if the response is different from one stimulus to another (two-sided two-sample test).

\begin{figure*}
	\begin{center}
		\includegraphics[scale=0.45]{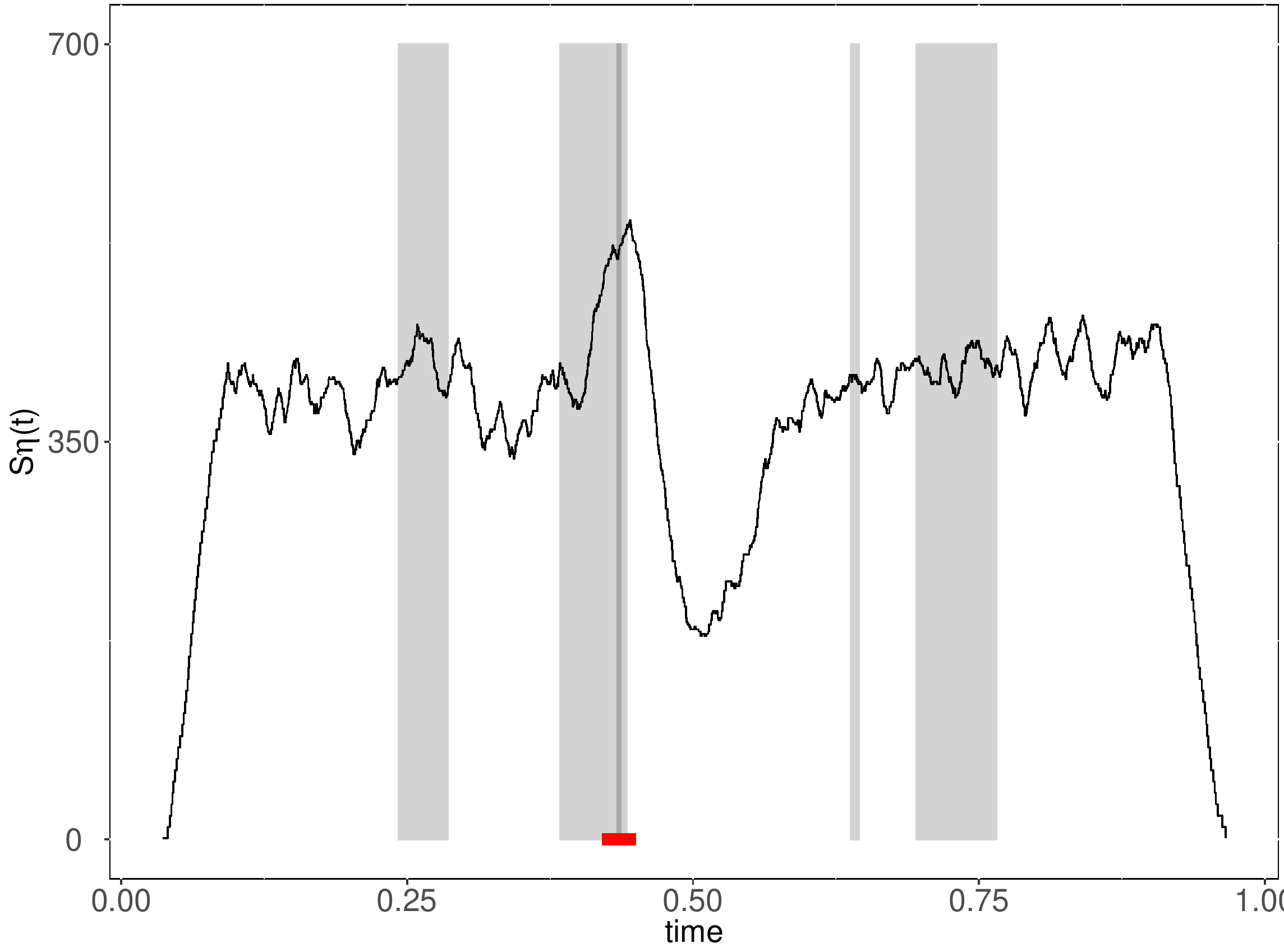}\\ 
		\includegraphics[scale=0.45]{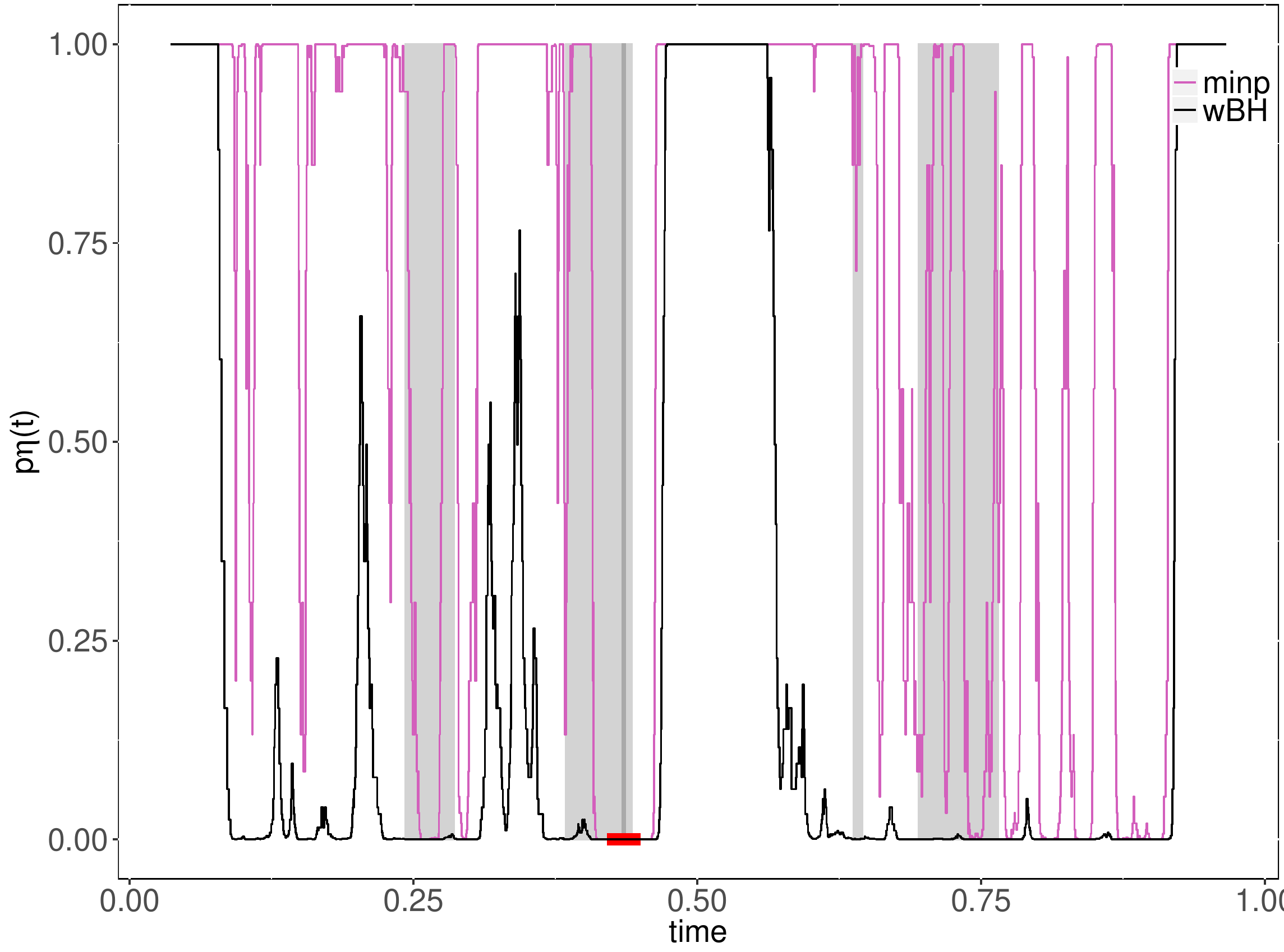}
	\end{center}
	\caption{\label{Fig:neuro2}Test of homogeneity for the spike trains of a cockroach neuron (neuron 2) excited by citronellal at $[0.42,0.45]$ (normalized time, red  band).  Top: count statistic $S_\eta(t)$ with $\eta=1/20$. Light-grey rectangles stand for intervals $\widehat{\mathcal{I}}_1^\eta$ on which the homogeneity hypothesis is rejected by the wBH procedure. The dark grey rectangle corresponds to the interval $\widehat{\mathcal{I}}_1^\eta$ that is rejected by the \minp procedure $(\alpha=5\%)$. Bottom: \pvalue process adjusted with the \minp or the wBH procedures.}
\end{figure*}

\begin{figure*}
	\begin{center}
		\includegraphics[scale=0.5]{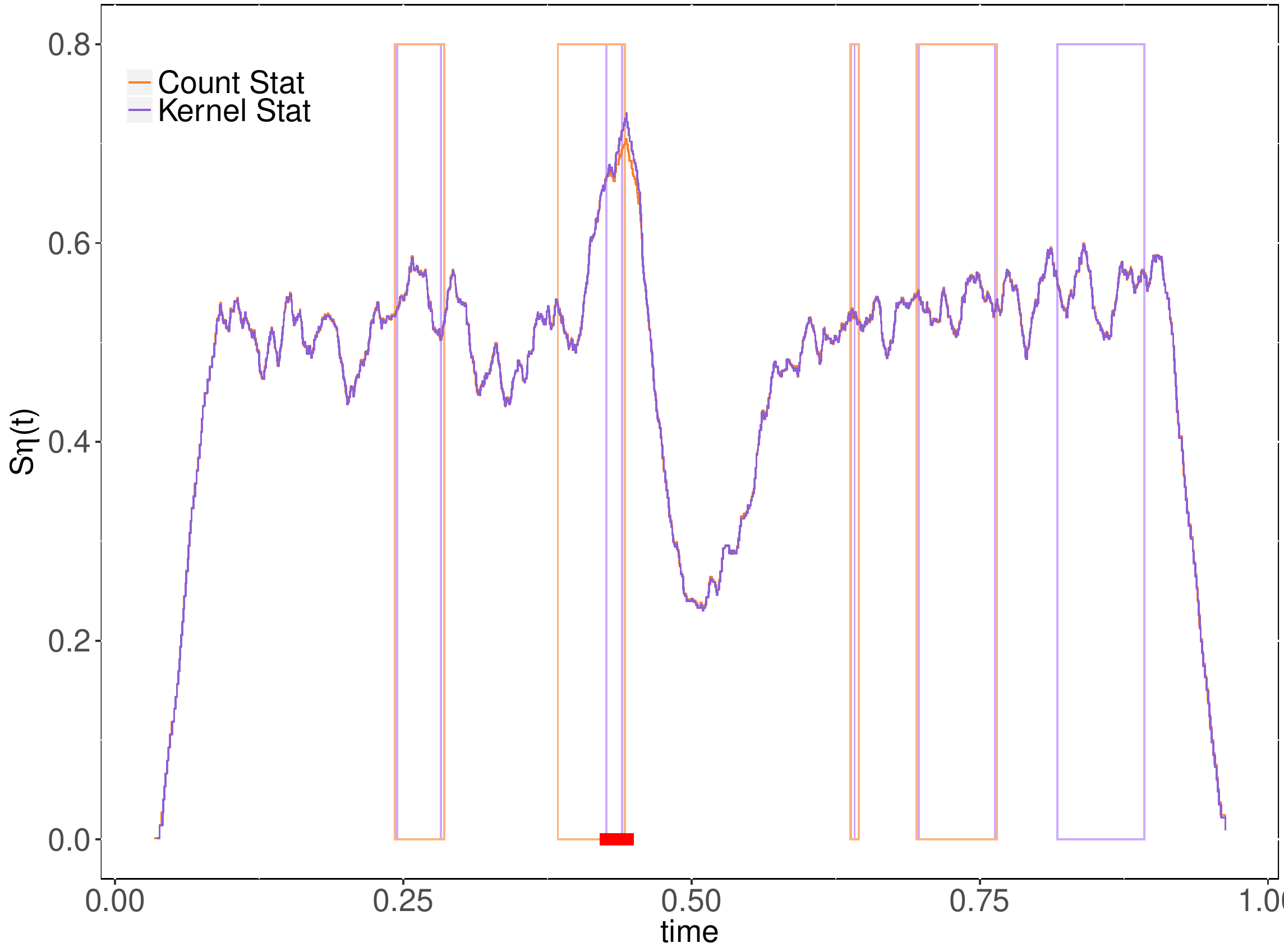}
	\end{center}
	\caption{\label{Fig:neuro2-comparison}Test of homogeneity ($\eta=1/20$) for the spike trains of a cockroach neuron (neuron 2) excited by citronellal at $[0.42,0.45]$ (normalized time, red vertical arrows). Rectangles stand for intervals $\widehat{\mathcal{I}}_1^\eta$ on which the homogeneity hypothesis is rejected by the wBH procedure $(\alpha=5\%)$.} 
\end{figure*}

\begin{figure*}
	\begin{center}
		\includegraphics[scale=0.45]{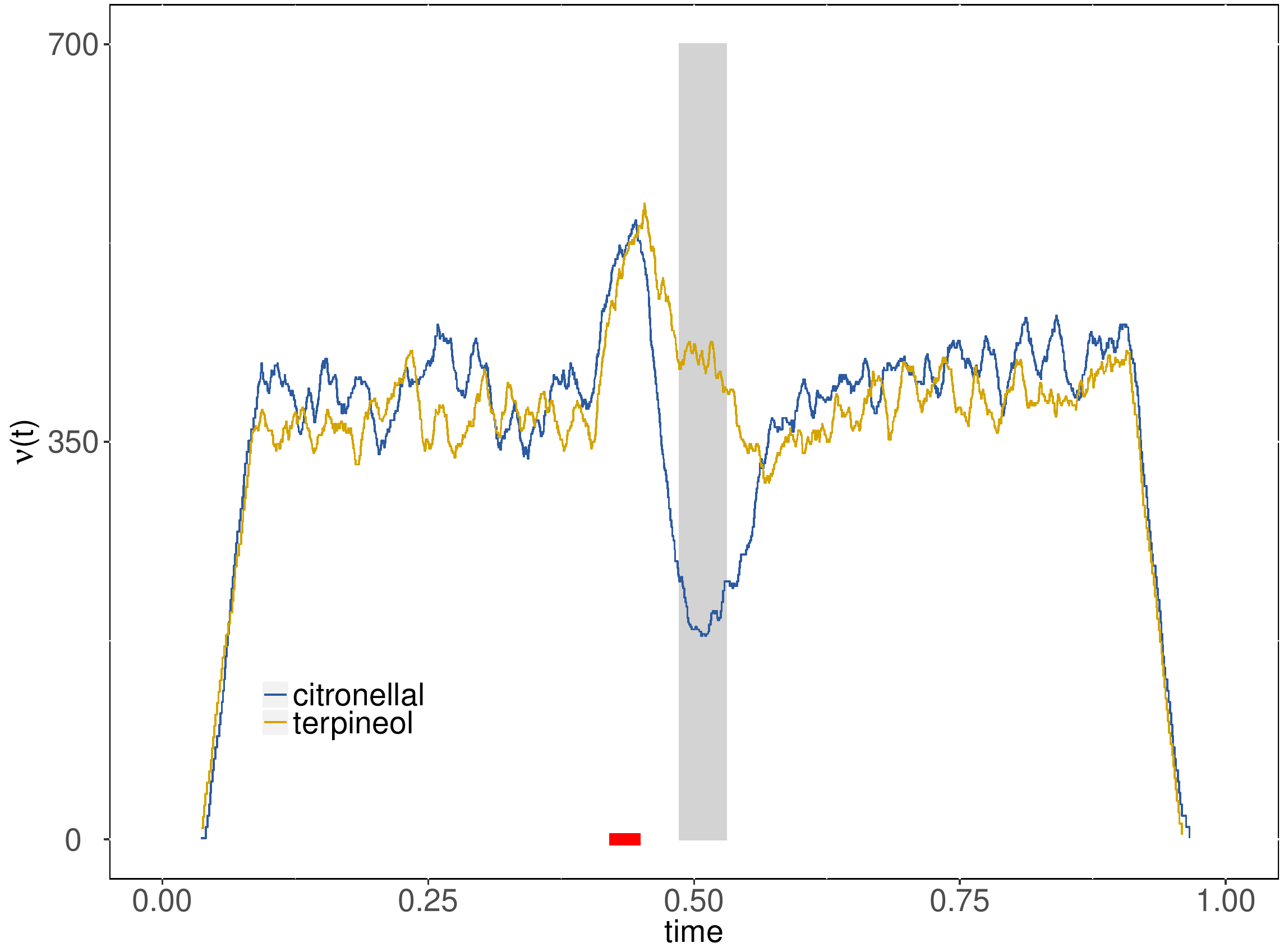}\\ 
		\includegraphics[scale=0.45]{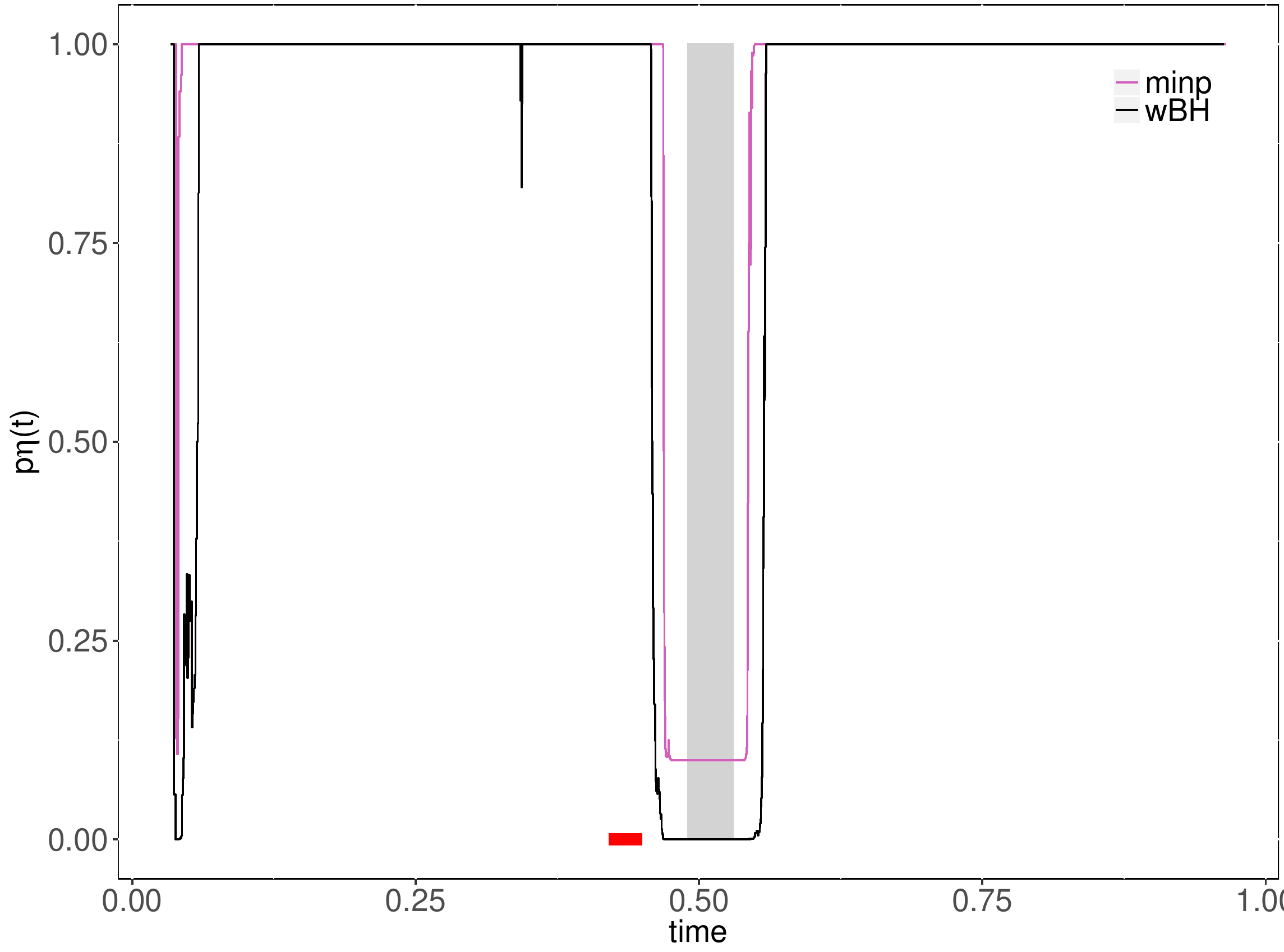}
	\end{center}
	\caption{\label{Fig:neuro2-twosample} Two-sample test for the spike trains of a cockroach neuron (neuron 2) excited by citronellal at $[0.42,0.45]$ (normalized time, red vertical arrows). Top: count statistic $S_\eta(t)$ with $\eta=1/20$. The dark grey rectangle corresponds to the interval $\widehat{\mathcal{I}}_1^\eta$ that is rejected by the wBH procedure $(\alpha=5\%)$. Bottom: \pvalue process adjusted with the \minp or the wBH procedures.}
\end{figure*}

The neuronal activity starts to be recorded at time 0, and the citronellal is presented between 0.42 and 0.45 (normalized time, Fig. \ref{Fig:neuro2}). Using the count statistic and the \minp procedure, our method identifies a neuronal response only when the neuron is stimulated (Fig. \ref{Fig:neuro2}, dark-grey rectangle), but not after the stimulus, whereas the weighted BH procedure identifies two firing rate changes posterior to the stimulus. Interestingly, the weighted BH procedure also identifies a deviation before the stimulus, which  may correspond to habituation and therefore anticipation of the stimulus by the animal (see for a similar phenomenon with synchrony instead of firing rates modifications \citep{MTGAUE}). What is remarkable is that the Gauss Kernel statistic globally identifies the same intervals of changes when compared with the count statistic (Fig. \ref{Fig:neuro2-comparison}), with the notable difference that the instantaneous response to the stimulus is located more precisely, which illustrates the higher specificity of the kernel-based statistics. Moreover, the kernel detects an additional interval of change in the firing rate. When studying the differential response of the same neuron to different stimuli (citronellal, terpineol) the differential change of fire rate occurs just after the stimulus (Fig. \ref{Fig:neuro2-twosample}). This difference is identified by the weighted BH procedure, but not by 
the \minp, which lacks of power, as already documented.

\subsection{\label{real} Differential analysis of replication origins between cell lines}

Next Generation Sequencing technologies (NGS) are based on the massive parallel sequencing of short DNA fragments, which have given rise to a new area in the field of high throughput biology. The applications of such technologies are many, and we focus on the particular cases where NGS allow the fine mapping of genomic features along the genome, like transcription factors (using the ChIP-Seq technology), or copy number variations (CNV-Seq) for instance. When applied to the field of replication, NGS technologies have allowed the fine-scale mapping of replication origins along the human genome. Briefly, replication is the mechanism by which genomes are duplicated into two exact copies. Genomic stability is under the control of a spatio-temporal program that orchestrates both the positioning and the timing of firing 100,000 replication starting points, also called replication origins \citep{PCA14}. This fine mapping has allowed us to identify potential regulators of the replication dynamics, and the comparative analysis of replication origins maps can help to identify a core of origins that are common to all human cells, compared with cell-type specific origins more related to cell differentiation. Moreover, pathologies like cancer being linked to replication instabilities, comparing replication origins in cancer cells vs. non-cancer cells could be promising. 

The chromosomal organization of replication origins can easily be modeled by a heterogeneous Poisson process of unknown intensity, and we consider the comparison of origins between cancer cells (Hela) and embryonic stem cells (iPS). Since the total number of origins differ from one condition to another and that globally there is no real difference in shape between both intensities except at some particular positions of interests, we need to account for the difference in the total number of occurrences per condition. To proceed, we modify the conditional distribution of the marks under the null, which becomes conditional to the total (random) number of points coming from condition A and B respectively:
$$
\varepsilon_T | N, \lambda_A, \lambda_B \underset{\mathcal{H}_0}{\sim} 2 \mathcal{B} \left(\frac{\lambda_A}{\lambda_A+\lambda_B}\right)-1,
$$
with $\lambda_A=\int_{0}^1\nu_A(t)dt$ and $\lambda_B=\int_{0}^1\nu_B(t)dt$. Since in practice $\lambda_A$ and $\lambda_B$ are unknown, they are replaced by the total number of points in $N_A$ and $N_B$. This procedure resumes to testing the equality of density functions instead of intensities.

\begin{figure*}
	\begin{center}
		\includegraphics[scale=0.45]{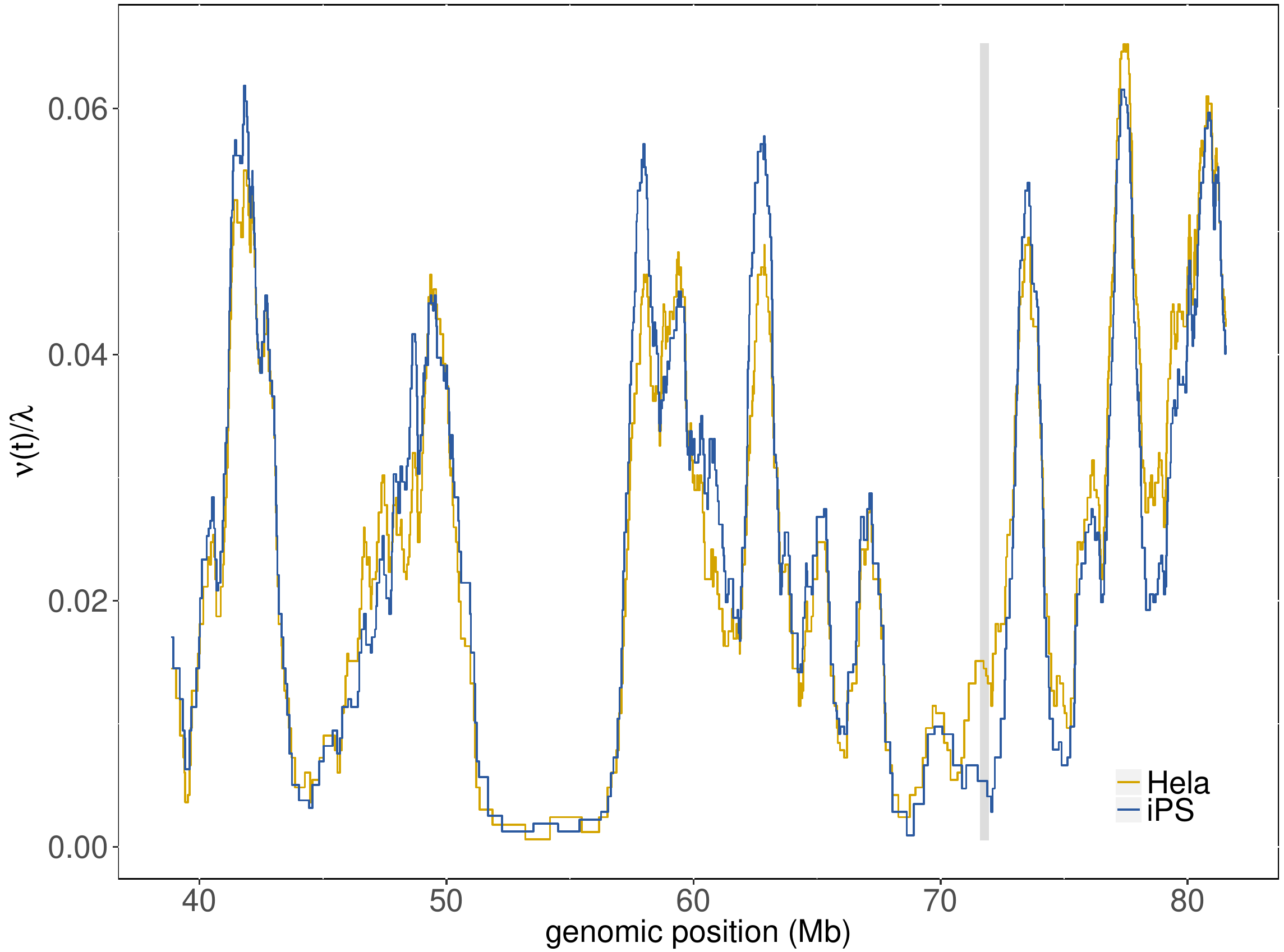} \\
		\includegraphics[scale=0.45]{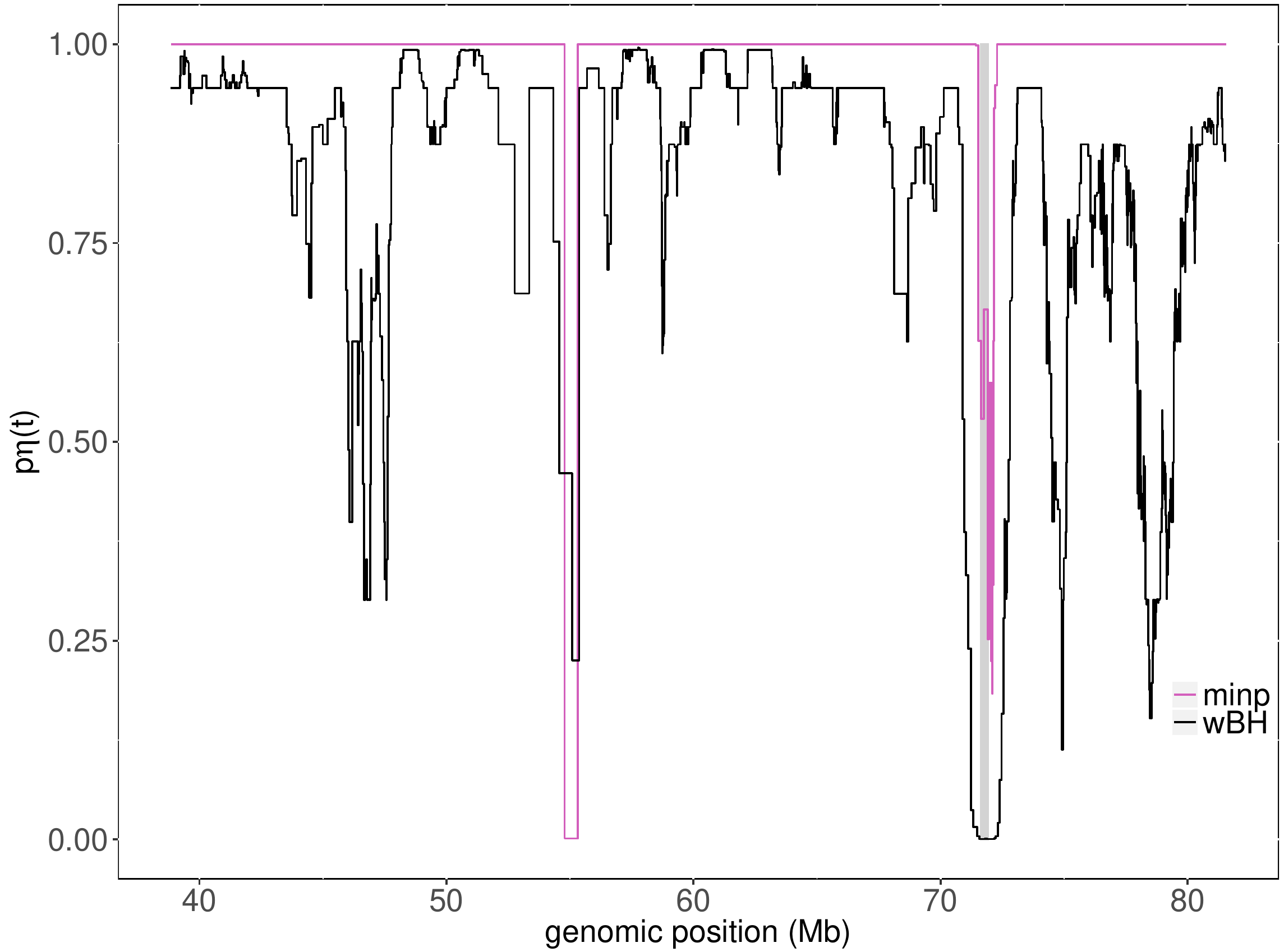}
	\end{center}
	\caption{\label{Fig:origins}Top: density of replication origins along chromosome 16-q, for Hela and iPS cells. The grey rectangle corresponds to the interval ($[71.64, 71.91]$ in Megabases) on which the equality hypothesis is rejected at level $\alpha=0.1$ using the count statistic, after adjustment by the wBH procedure. The Gaussian kernel statistics gives the same result. Bottom: \pvalue process adjusted by the wBH procedure (black), and the \minp procedure (purple). The \minp procedure does not reject any region. The size of the window corresponds to windows of 1 Mb.}
\end{figure*}

As expected, origins densities show strong variations along chromosomes, but those variations are globally comparable between cell lines (Fig. \ref{Fig:origins}). One region shows significantly more origins in Hela cells (cancer cells), that is detected by both the count and the Gaussian kernel statistic after the continuous wBH procedure. The \minp adjustment does not allow the detection of this region at the same nominal level, but the variations of the adjusted \pvalue process indicate a strong difference of densities between cell lines. This region corresponds to a gene-rich region that needs investigation for further biological characterization. More and more cell lines have been investigated to determine the core of replication origins that are common to any human cells, and the set of origins that are cell-type specific \citep{BBL12}. This context offers a nice perspective of the present work, by generalized our two-sample test to the more general framework of a $k$-sample (or ANOVA-like) test.

\section{Conclusion}

We proposed a new framework for testing Poisson processes intensities based on continuous testing. This framework appears very powerful, from both the theoretical and the practical points of views. It allows the complete definition of the testing procedure, based on local null hypothesis, which provides a clear definition of false positive windows, that was lacking in existing works. We introduced the \pvalue process that becomes a stochastic process that needs to be adjusted to account for joint error rates. Contrary to existing approaches, continuous testing avoid the arbitrary discretization of this process, which is coherent with the continuous nature of the underlying model. The continuous framework especially provides a rigorous definition of error rates in continuous time, and we proposed two procedures to control the FWER and the FDR. In addition to theoretical controls, we show the empirical properties of our procedure, that can be easily implemented, and is available in the form of the \texttt{contest} \texttt{R}-package. The application on experimental data inspires extensions of our framework, that could be adapted to the testing of more than 2 conditions, or to other tests like local independence tests that are essential in Neurosciences \citep{ABF15}.  Calibrating the window size is certainly one of the major perspective of this work. While the subject has been extensively investigated in the context of model estimation, only a few works considered the question in the testing framework \citep{GG08}. Consequently, one challenge ahead will be to determine if the continuous testing framework would be appropriate to derive a calibration method to automatically select the resolution of the test from the data. Other future extensions involve either spatial aspects or more general point processes from a theoretical point of view.

\section*{Acknowledgment} We are grateful to Anne-Laure Foug\`eres for fruitful discussions on the subject, to Matthieu Lerasle for sharing with us the trick of the double Monte-Carlo step, and to Ivan Bardet for his work on the \cite{CZ07} paper as an intern. This research was partly supported by the french Agence Nationale de la Recherche (ANR 2011-BS01-010-01 Calibration project, ANR-11-BINF-0001-06 ABS4NGS project, ANR-15-CE12-0004-02 OriMolMech project and ANR-16-CE40-0019 SansSouci project). This work was performed using the computing facilities of the computing center LBBE/PRABI.

\bibliographystyle{chicago}
\bibliography{PRR17}

\newpage
\appendix

\section{Supplementary material}\label{appendixA}

\subsection{Step-down improvement} \label{sec:SD}

Let us first introduce some notation.
For any fixed subset $\mathcal{C}$ of $\mathcal{X}_\eta$,  $F_{\theta_0,\Nsf}^{\mathcal{C}}$ is the conditional c.d.f. given $\Nsf$ of $\inf_{x \in \mathcal{C}} p_\eta(x)$. Then we define 
$$\mathcal{N}(\mathcal{C})= \{ x \in \mathcal{X}_\eta , F_{\theta_0,\Nsf}^{\mathcal{C}} (p_\eta(x))\leq \alpha\}.$$
The theoretical step-down algorithm can then be derived as follows:
\begin{itemize}
	\item Step 0 : compute $\mathcal{R}^0=\emptyset$ given by \eqref{FWERprocedureq}.
	\item Step $j\geq 1$: compute 
	$\mathcal{R}^j=\mathcal{N}\left((\mathcal{R}^{j-1})^c\right)$
\end{itemize}
The step-down rejection set $\mathcal{R}^{step-down}_\eta$ is then defined as the limit of $\mathcal{R}^j$ when $j$ tends to infinity. Note that
$\mathcal{R}^1=\mathcal{R}_\eta^{\m}$ is in fact given by \eqref{FWERprocedureq} and that the sequence of $\mathcal{R}^j$ is increasing.

The fact that this algorithm ends almost surely in a finite number of steps is just an easy consequence of the fact that for a given realization, there is only a finite number of possible values for the \pvalue process, which is in fact upper bounded by the size of the partition $\tau$. Indeed, at each round, $\mathcal{R}^j$ cannot  decrease  and if it increases, it is by absorbing sets of the type $\{x \in \mathcal{X}_\eta, p_\eta(x)=a\}$ for some $a$.

One can prove that such an algorithm guarantees  a controlled FWER following the ideas of \citep{GoemanSolari}. More precisely if one defines the event 
$$\mathcal{E}=\{ J_0^\eta \cap \mathcal{N}(J_0^\eta)=\emptyset\},$$
one has by $\left(\mathcal{P}\right)$ (see Section \ref{gendef}) that
\begin{eqnarray*}
	P_{\theta,\lab}(\mathcal{E}^c)& = & P_{\theta,\lab}(\exists x \in J_0^\eta, F_{\theta_0,\Nsf}^{J_0^\eta} (p_\eta(x)) \leq \alpha)\\
	& = & P_{\theta_0,\lab}(\exists x \in J_0^\eta, F_{\theta_0,\Nsf}^{J_0^\eta} (p_\eta(x)) \leq \alpha)\\
	& = & E_{\lab}\left[ P_{\theta_0}\left( F_{\theta_0,\Nsf}^{J_0^\eta} (\inf_{x\in J_0^\eta} p_\eta(x)) \leq \alpha\: \bigg|\: \Nsf\right)\right]\leq \alpha.
	%\\&\leq & \alpha
\end{eqnarray*}
The last point comes from the fact that $F_{\theta_0,\Nsf}^{J_0^\eta} (\inf_{x\in J_0^\eta} p_\eta(x))$ can be seen as the \pvalue of the test based on the statistics $- \inf_{x\in J_0^\eta} p_\eta(x)$ and described for instance in Lemma 1 of \citep{FWSR}, which implies that \eqref{equdefpvalues} holds even for the \pvalue $F_{\theta_0,\Nsf}^{J_0^\eta} (\inf_{x\in J_0^\eta} p_\eta(x))$. This is the single-step property of \citep{GoemanSolari}.

\noindent For the monotony property of \citep{GoemanSolari}, remark that if 
$\mathcal{C}\subset \mathcal{C'}$ then $F_{\theta_0,\Nsf}^{\mathcal{C}}\leq  F_{\theta_0,\Nsf}^{\mathcal{C'}}$ and $\mathcal{N}(\mathcal{C}')\subset\mathcal{N}(\mathcal{C})$. Hence for all $j$, 
if $J_0^\eta \subset (\mathcal{R}^j)^c$, $R^{j+1}=\mathcal{N}([\mathcal{R}^j]^c)\subset \mathcal{N}(J_0^\eta)$.

Now, on $\mathcal{E}$, $\mathcal{N}(J_0^\eta)\cap J_0^\eta=\emptyset$ and therefore if $J_0^\eta \subset (\mathcal{R}^j)^c$, $J_0^\eta \subset (\mathcal{R}^{j+1})^c$ on the same event. 
Since $J_0^\eta \subset (\mathcal{R}^0)^c=\mathcal{X}_\eta$, one has with probability larger than $1-\alpha$ that $J_0^\eta\subset \lim_j \mathcal{R}^j= \mathcal{R}_\eta^{step-down},$ which means that $\FWER_{\theta,\lab}^\eta(\mathcal{R}_\eta^{step-down}) \leq \alpha.$

\subsection{Other kernel-based single test statistics}\label{sec:singletesting}

\subsubsection{Homogeneity test statistics}\label{sec:homstat}

\paragraph{Two sided test statistics.} The starting point is that, conditionally on the event $\{N([0,1])=n\}$, the points in $N$ form an $n$ i.i.d. sample with density $f = 1+\theta$. Hence, for any translation invariant  kernel $K_h:[0,1]\to \R$ of bandwith $h$, we have 
$$\widehat{f}_h(s) = \frac{1}{n}\sum_{T\in N\cap I_\eta(x)} K_h(s-T),$$ is an unbiased consistent estimate (when $n$ tends to infinity) of 
$$f_h(s)=\int_{I_\eta(x)} K_h(s-t) f(t) dt.$$
Consequently, the corresponding $U$-statistics given by
\begin{equation}
\frac{1}{n(n-1)} \sum_{\substack{T' \in N\cap I_\eta(x)\\T'\not = T}} K_h(T-T'),
\end{equation}
is an unbiased estimate of:
$$\int \int_{I_\eta(x)^2} K_h(s-t) f(t)f(s) ds \underset{h \rightarrow 0}{\longrightarrow}\|f\|^2_{I_\eta(x)},$$
under standard conditions on the kernel $K_h$. Actually, in a testing framework, we do not want to estimate $\|f\|_{I_\eta(x)}^2$ \textit{stricto sensu}: here, we do not need to consider small values for $h$, we even recommend to take $h=\eta$ when it is unknown. This choice would not be appropriate for estimation, but seems sufficient for testing. Moreover if $\mathcal{H}_0\left\{I_\eta(x)\right\}$ holds, $f=1$ on $I_\eta(x)$ and $\|f\|^2_{I_\eta(x)}=\eta$. So we propose the following statistics
\begin{equation*}
S_\eta(x) = \left|  \frac{1}{n(n-1)} \sum_{\substack{T' \in N\cap I_\eta(x)\\T'\not = T}} K_h(T-T') -\eta\right|,
\end{equation*}
and the local null hypothesis is rejected for high values of $S_\eta(x)$.

\paragraph{One-sided test statistic.} In this case we focus on what is happening when $f>1$. Hence the target quantity to estimate is $$\int \int_{I_\eta(x)^2} f(s) \max\big(f(s),1\big) ds.$$ Here we consider the following density estimator:
$$
\widehat{f}_h(s) = \frac{1}{(n-1)} \sum_{\substack{T' \in N\cap I_\eta(x)\\T'\not = s}} K_h\left(T'-s\right),$$
which leads to the following statistics:
\begin{equation*}
\label{Homo_onesided}
S_\eta(x)= \frac{1}{n} \sum_{ T \in N\cap I_\eta(x)} \max \left(\widehat{f}_h(T), \quad 1\right).
\end{equation*}
If there exists a $t$ in $I_\eta(x)$ such that $f(t)>1$ and if $f$ is smooth enough such that this holds on a neighborhood, $\E[\max(f(T),1)\ind{T\in I_\eta(x)}]$ should be larger than $\eta$, whereas its value should be smaller than $\eta$ if there are no such $t$. 

\subsubsection{Two-sample test statistics (one-sided)}\label{sec:twostat}

Applying a reasoning similar to above, the quantity $$\frac{1}{N([0,1])-1}\sum_{\substack{T' \in N\cap I_\eta(x)\\T'\not = T}} K_h(T-T') \eps_{T'},$$ is a good estimate of $\theta(T)\lab(T)$, for each $T\in N$. If one wants to reject only when there exists a $t \in I_\eta(x)$ such that $\theta(t)>0$, then one needs to take the positive part of the previous quantity and integrate it with respect to $T \in N\cap  I_\eta(x)$. We do not want to multiply again by $\eps(T)$ because we do not want to multiply  by $\theta$  twice. This leads to
\begin{equation*}
\label{TwoSample_onesided}
S_\eta(x)=\frac{1}{N([0,1])}\sum_{T \in N\cap  I_\eta(x)} \max\left( \frac{1}{N([0,1])-1}\sum_{\substack{T' \in N\cap I_\eta(x)\\T'\not = T}} K_h(T-T')  \eps_{T'}, \quad 0\right),
\end{equation*}
which estimates $\E_\lambda\big[\max(\theta(T),0)\ind{T\in I_\eta(x)}\big]$ for $T$ of density proportional to $\lab$.

\section{Proofs}\label{appendixB}

\subsection{Proof of Theorem~\ref{th:FWERSD}}\label{sec:proofFWERSD}

Computations and arguments similar to Section \ref{sec:SD} gives that
\begin{eqnarray*}
	\FWER_{\theta,\lab}^\eta({\mathcal{R}}_\eta^{\m}) &=& E_\lab \left[P_\theta\left(\exists x \in J_0^\eta, q_\eta(x) \leq \alpha\: |\: \Nsf \right)\right]\\
	&=&  E_\lab \left[P_\theta\left(F^{\m}_{\Nsf}\left(\inf_{x \in J_0^\eta} p_\eta(x)\right) \leq \alpha \:\bigg|\: \Nsf \right)\right]\\
	&=&  E_\lab \left[P_{\theta_0}\left(F^{\m}_{\Nsf}\left(\inf_{x \in J_0^\eta} p_\eta(x) \right)\leq \alpha \:\bigg|\: \Nsf \right)\right]\\
	&\leq &  E_\lab \left[P_{\theta_0}\left(F^{\m}_{\Nsf}\left(\inf_{x \in \mathcal{X}_\eta} p_\eta(x) \right)\leq \alpha \:\bigg|\: \Nsf \right)\right]\leq \alpha,
	%&\leq & \alpha,
\end{eqnarray*}
which concludes the proof.

\subsection{Proof of Theorem~\ref{th-FDR}}\label{sec:proofth-FDR}

We follow the methodology introduced in \cite{BDR2014} by applying their Theorem~4.1 in the "Poisson case", as discussed in Example~2.5 and Section~4.2.4 therein.
While the measurability conditions can be easily checked because the $p$-value process is c\`adl\`ag, the result will be proved if we show that the $p$-value process is finite-dimensional strong PRDS on $J_0^\eta$, as defined in \cite{BDR2014}.
For the homogeneity test,  Lemma~A.2 in \cite{BDR2014} shows that the $p$-value process is finite-dimensional strong PRDS on any subset, and thus also on $J_0^\eta$.
For the two-sample test, this condition takes the following form: for any 
$q\geq 1$, for any $(x_j)_{1\leq j\leq q}\in (\mtc{X}_\eta)^q$ such that $x_1\in J_0^\eta$, for any nonincreasing set $D\subset \mathbb{N}^q$, for any $n\geq 0$,
\begin{align}
\nonumber
&\P_{\theta,\lambda} \left(S_\eta(x_j))_{1\leq j \leq q}  \in D  \:|\: S_\eta(x_1)=n+1\right) \\
&\leq  \P_{\theta,\lambda} \left( (S_\eta(x_j))_{1\leq j \leq q}  \in D  \:|\: S_\eta(x_1)=n\right).\label{PRDS-Poisson} 
\end{align}

Above, a nonincreasing set is defined as a subset $D\subset \mathbb{N}^q$ such that for all $(m,m') \in \mathbb{N}^q$ with $ m_j\leq
m'_j$ for $ 1\leq j \leq q$, we have $m' \in D \Rightarrow m \in D$.

Now, to establish \eqref{PRDS-Poisson}, we work {\it conditionally} on the joint Poisson process $N$. In that case, the random part of $S_\eta(x)=\sum_{T \in N\cap I_\eta(x)} (2 \eps_T-1)$ is only carried by the set of marks  $(\eps_T)_{T\in N}$. Let $b_T=2 \eps_T-1$ for $T\in N$ and 
$$
D'=\left\{(b_T)_{T\in N} \in \{0,1\}^{N}\::\:  \left(\sum_{T \in N\cap I_\eta(x_j)} b_T \right)_{1\leq j \leq q}  \in  D  \right\}.
$$
Since $D$ is a nonincreasing set, so is $D'$ and thus there exists some subset $\mtc{T} \subset N$ such that $D'=\{ (b_T)_{T\in N} \in \{0,1\}^{N}\telque \forall T \in \mtc{T}, b_T=0\}$. Furthermore, since 
$(b_T)_{T\in N \cap I_\eta(x_1)}$ is independent of $(b_T)_{T\in N \backslash I_\eta(x_1)}$, we have
\begin{align}
&\P_{\theta}\left( (S_\eta(x_j))_{1\leq j \leq q}  \in D  \:|\: S_\eta(x_1)=n,  N\right) \nonumber\\
&= \P_\theta \left( \forall T \in \mtc{T}, b_T=0  \:\bigg|\: \sum_{T \in N\cap I_\eta(x_1)} b_T = n,  N\right) \nonumber\\
&=  \P_\theta \left(\forall T \in \mtc{T}\backslash I_\eta(x_1) , b_T=0  \right) \\
&\times \P_\theta \left(\forall T \in \mtc{T} \cap I_\eta(x_1), b_T=0  \:\bigg|\: \sum_{T \in N\cap I_\eta(x_1)} b_T = n,  N\right) \nonumber\\
&=  \P_\theta \left(\forall T \in \mtc{T}\backslash I_\eta(x_1) , b_T=0  \right) \times  \binom{U -V}{n} / \binom{U}{n},\label{equintermth3}
\end{align}
by denoting $U=\#N\cap I_\eta(x_1)$, $V=\#\mtc{T}\cap I_\eta(x_1)$ and by an elementary combinatorial argument using that $b_T$, $T\in N\cap I_\eta(x_1)$, are i.i.d. $\mtc{B}(1/2)$ variables.
Finally, since $\binom{U -V}{n} / \binom{U}{n}= \binom{U -n}{V} / \binom{U}{V}$, the right-hand-side of \eqref{equintermth3} is nonincreasing w.r.t. $n$ and the relation \eqref{PRDS-Poisson} is proved.

\subsection{Proof of Theorem~\ref{th:pvaltwoMC}}\label{sec:proofthFWERSDhat}

Thanks to  $\left(\mathcal{P}\right)$  (see Section \ref{gendef}) that is used for the observed sample and the fact that the distribution of the resampled $\varepsilon^b$ for $b=1,...,B$, is fixed whatever $\theta$, we have that
\begin{align}
\FWER_{\theta,\lab}^\eta(\hat{\mathcal{R}}_\eta^{\m})&=\E_\lab \left[P_\theta\left(\exists x \in J_0^\eta, \widehat{q}_\eta(x)\leq \alpha\:\bigg|\: \Nsf \right)\right]\nonumber\\
&= \E_\lab \left[P_\theta\left(\frac{1}{B+1} \left(1+\sum_{b=1}^{B}\ind{m^{b} \leq \inf_{x\in J_0^\eta} \hat{p}_\eta(x)}\right)\leq \alpha \:\bigg|\: \Nsf \right) \right]\nonumber\\
&=   \E_\lab \left[P_{\theta_0}\left(\frac{1}{B+1} \left(1+\sum_{b=1}^{B}\ind{m^{b} \leq \inf_{x\in J_0^\eta} \hat{p}_\eta(x)}\right)\leq \alpha\:\bigg|\: \Nsf \right)\right]\nonumber\\
&\leq  \E_\lab \left[P_{\theta_0}\left(\frac{1}{B+1} \left(1+\sum_{b=1}^{B}\ind{m^{b} \leq \inf_{x\in \mathcal{X}_\eta} \hat{p}_\eta(x)}\right)\leq \alpha \:\bigg|\: \Nsf \right)\right]. \nonumber
\end{align}
As a consequence, by using the notation $m^0$, we have
\begin{align*}
\FWER_{\theta,\lab}^\eta(\hat{\mathcal{R}}_\eta^{\m})
\leq  \E_\lab \left[P_{\theta_0}\left(\frac{1}{B+1} \left(1+\sum_{b=1}^{B}\ind{m^{b} \leq m^0} \right)\leq \alpha \:\bigg|\: \Nsf \right)\right].
\end{align*}
Since, under $P_{\theta_0}$, the  random variables $\eps^0,...,\eps^B$ are i.i.d. Rademacher, the process vector $([\widehat{p}^{b}_{\eta}(x)]_{x\in \mathcal{X}_\eta})_{b=0,...,B}$ is exchangeable (under $P_{\theta_0}$ and cond. on  $\Nsf$), which in turn implies that the vector $(m^b)_{b=0,...,B}$ is exchangeable (under $P_{\theta_0}$ and cond. on  $\Nsf$). 
Therefore, applying Lemma 1 of \citep{RomanoWolf}, we obtain
$$\P_{\theta_0}\left( \frac{1}{B+1} \left(1+\sum_{b=1}^{B}\ind{m^{b} \leq m^0}\right)\leq \alpha \:\bigg|\: \Nsf \right) \leq \alpha,$$
and it remains to integrate in $\Nsf$ to conclude.

\begin{figure*}
	\begin{center}	
		\begin{tabular}{cc}
			\includegraphics[scale=0.35]{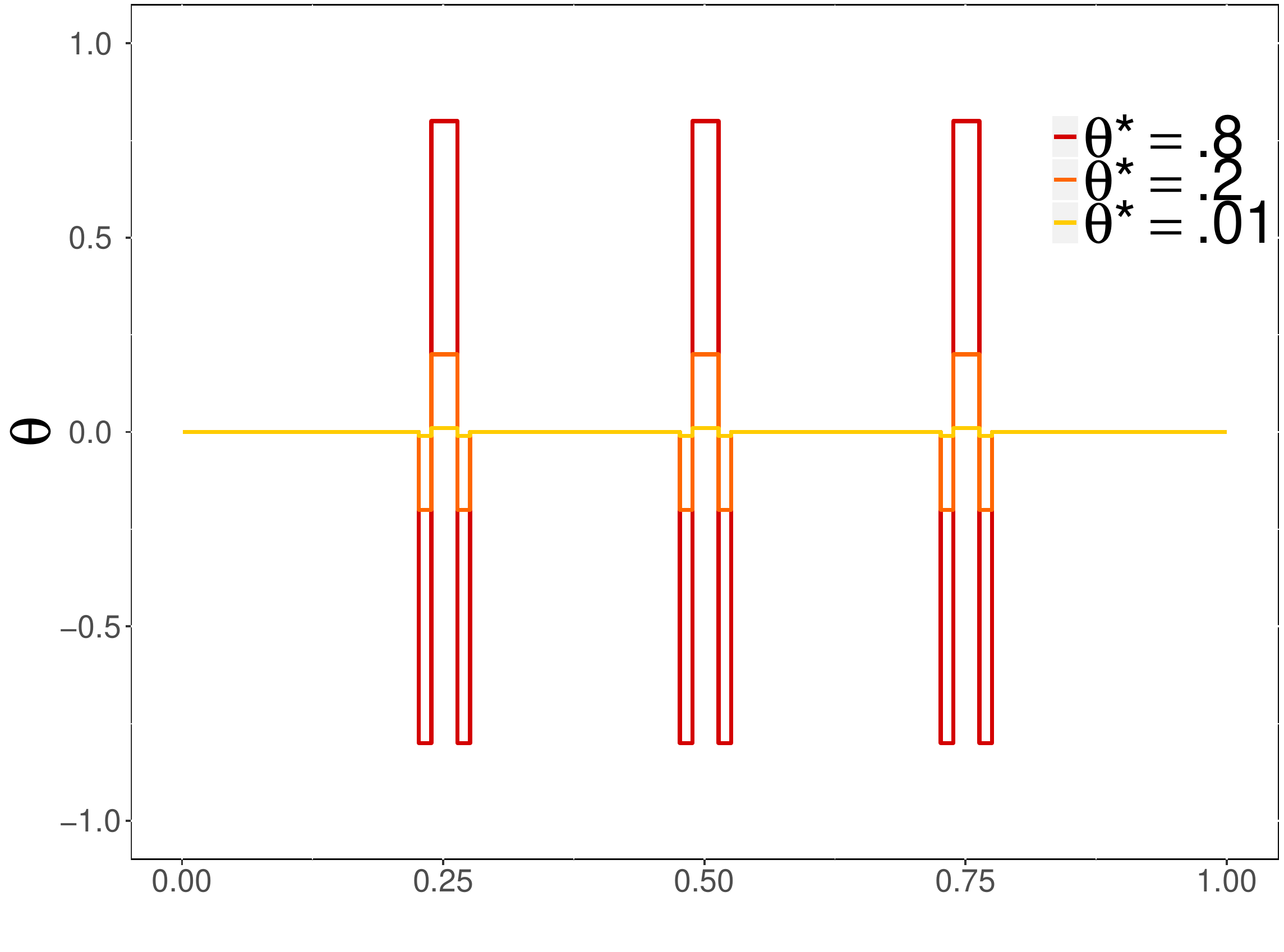} 
		\end{tabular}
	\end{center}
	\caption{\label{Fig:theta}Simulation setting for the homogeneity case. Signal function $\theta(.)$ that equals $0$ under the null hypothesis, and $\theta^*$ elsewhere.}
\end{figure*}

\end{document}